\documentclass[smallcondensed]{svjour3}
\smartqed  
\usepackage{graphicx,amssymb,float,natbib,multirow}
\begin{document}

\title{A Mechanism for Discovering Semantic Relationships among Agent Communication Protocols
}


\author{Idoia Berges \and Jes\'us Berm\'udez \and Alfredo Go\~ni 
\and Arantza Illarramendi 
}


\institute{I. Berges \at
              University of the Basque Country, Paseo Manuel de Lardizabal, 1, Donostia-San Sebasti\'an, Spain \\
              Tel.: +34-943015109\\
              \email{idoia-berges@ikasle.ehu.es}           
           \and
           J. Berm\'udez\\
\email{jesus.bermudez@ehu.es}\\
		 \\
A. Go\~ni\\ \email{alfredo@ehu.es}\\
	 \\
A. Illarramendi\\ \email{a.illarramendi@ehu.es}
}

\date{Received: date / Accepted: date}

\maketitle

\begin{abstract}
One relevant aspect in the development of the Semantic Web framework is the achievement of a real inter-agents communication capability at the semantic level. Agents should be able to communicate with each other freely using different communication protocols, constituted by communication acts.

For that scenario, we introduce in this paper an efficient mechanism presenting 
the following main features:
\begin{itemize}
\item 
It promotes the description of the communication acts of protocols as classes that belong to a communication acts ontology, and associates to those acts a social commitment semantics 
formalized through predicates in the Event Calculus.

\item
It is sustained on the idea that different protocols can be compared semantically by looking to the set of fluents associated to each branch of the protocols. Those sets are generated using Semantic Web technology rules.

\item
It discovers the following types of protocol relationships: equivalence, specialization, restriction, prefix, suffix, infix and complement\_to\_infix.
\end{itemize}

\end{abstract}

\section{Introduction}
\label{intro}

In the scenario promoted by the emerging Web, administrators of existing Information Systems distributed along the Internet are encouraged to provide the functionalities of those systems, either through agents that represent them or through Web Services.
The underlying idea is to get real interoperation  among those Information Systems in order to enlarge the benefits that users can get from the Web by increasing machines' processable tasks.

Although agent technology and Web Services technology have been developed in separate ways, there exists a recent
work \citep{GLMS07} which tries to consolidate their approaches into a common
 specification describing how to seamlessly interconnect
 FIPA compliant agent systems \citep{FIPA-ACL}
with W3C compliant Web Services.
The purpose of specifying an infrastructure for integrating
these two technologies is to provide a common means of
allowing each to discover and invoke instances of the other. Considering the previous approach, in the rest of this paper we will only concentrate on aspects of inter-agent communication. 

In general, communication among agents is based on the interchange of messages, which in this context are also known as communication acts. However, since Information Systems are developed independently, they incorporate different classes of communication acts as their Agent Communication Language  (\textsc{acl}) to the point that they do not understand each other. Moreover, protocols play an integral role in agents communication. A protocol specifies the rules of interaction between agents by restricting the range of allowed follow-up communication acts for each agent at any stage during a communicative interaction. 

The importance of using communication protocols is widely recognized. However, that does not mean that all the agents should use the same universal protocols. In our opinion different communication protocols should coexist and every agent should have the opportunity of selecting the one that better fulfills its needs. Thus, one possible procedure that the administrators of the Information Systems could follow in order to implement the agents that will represent their systems is the following: They should first select, from a repository of standard protocols (more than one repositories may exist), those protocols that best suit the goals of their agents. Sometimes a single protocol will be enough, while other times it will be necessary to design a protocol as a composition of other protocols. Then, the selected protocols may be customized before they are embedded into the agents. This flexibility at the time of choosing and customizing communication protocols has a drawback: it is likely that two protocols that have been designed for the same goal have different structure. Therefore, a reasoning process over the protocols embedded in the agents will be necessary to discover relationships -such as equivalence and restriction- between them.  

In this paper we present a mechanism that discovers semantic relationships 
among protocols focusing on three main aspects: 1. The semantic representation of the main elements of the protocols, that is, of communication acts; 2. The semantic representation of the protocols' branches using Semantic Web Technology; and 3. The definitions of a set of semantic relationships between protocols. 

Concerning the first point, communication acts that constitute protocols  are described as classes that belong to a communication acts ontology, which we have developed, called \textsc{CommOnt} (see more details about the ontology 
in~\citep{IS2007}). 
The use of that ontology 
favours both the explicit representation of the meaning of the 
communication acts and the customization of existing standard protocols by allowing the use of particular communication acts that can be defined as specializations of existing standard communication acts. 
We have adopted the so called \textit{social approach} \citep{Singh98,Singh00}
for expressing the  
intended semantics of those communication acts.
According to the social approach, when agents interact they become 
involved in social commitments or obligations to each other. 

With respect to the second point, one fundamental step of the mechanism is to extract the semantics associated to each protocol's branches. This step is sustained on the idea that the semantics of each branch is represented by a set of predicates generated after the whole branch is analyzed. During this step Semantic Web technology rules are used.

Finally, a semantically founded set of basic relationships between protocols is defined, which later can be composed to express more complex relationships. Our mechanism is able to discover those complex semantic relationships taking into account the previously mentioned protocol analysis.

 In summary, the main contributions of the mechanism presented in this paper are:
\begin{itemize}
\item It favours a flexible interoperation among agents of heterogeneous Information Systems that use different communication protocols and therefore avoids the need for prior agreement on how the messages are interchanged.

\item It facilitates the customization of standard communication protocols by means of specialized communication acts that belong to specific
\textsc{acl} of Information Systems. The particular communication acts are described in an ontology.
\item It provides a basis to reason about protocol relationships founded on the following basic relations: equivalence, specialization, restriction, prefix, suffix, infix and complement\_to\_infix. Moreover, notice that our approach allows the comparison between protocols in terms of the 
intended semantics of communication acts that appear in those protocols.

\end{itemize}

The rest of the paper is organized as follows: Section~\ref{pilares} provides 
background on the communication ontology, which contains terms corresponding to 
communication acts that appear in the protocols, and on the semantics associated to those acts. Section~\ref{description} explains how protocols are described in our proposal and introduces our algorithm that makes a semantic analysis of protocols. Section~\ref{ProtocolRelationships} presents the formal definitions of those relationships. In Section~\ref{Example} there is a scenario where the proposed mechanism is applied. Section~\ref{trabajos} discusses different related works, and the conclusions appear in the last section.

\section {Representation of the semantics of communication acts}\label{pilares}
\noindent Among the different models proposed for representing protocols one which stands out is that of State Transition Systems (STS).
\begin{definition}
A \textit{State Transition System} is a tuple ($S$, $s_{0}$, $L$, $T$, $F$), where 
$S$ is a finite set of states, $s_{0} \in S$ is the initial state, $L$ is a finite set 
of labels, $T\subseteq S\times L\times S$ is a set of transitions and $F\subseteq S$ 
is a set of final states. 
\end{definition}
In our proposal we use STSs where transitions are labeled with communication act classes  
described in a communication acts ontology called \textsc{CommOnt}. 
That is to say, the set of labels $L$ is a set of class names taken from that ontology. 
\textsc{CommOnt} is a \textsc{OWL-DL} ontology, therefore its term descriptions are founded on 
a Description Logic (DL). DLs are well designed for expressing conceptual knowledge and 
representing static structural knowledge. However, protocols exhibit a dynamic aspect 
due to the effect of communication acts. Thus, we have decided to represent the effects of 
actions through predicates in the Event Calculus.

In a nutshell, the \textsc{CommOnt} ontology describes structural and hierarchical aspects of 
communication acts and a knowledge base of Event Calculus predicates specifies the effects 
of communication acts. Terms that appear in the Event Calculus predicates come from names 
of classes and properties in \textsc{CommOnt}. 

Reasoning over the effects of communication acts is achieved through an encoding of event 
calculus predicates and rules into SWRL (Semantic Web Rule Language). 
SWRL is an extension of \textsc{OWL-DL} axioms to include Horn-like rules. Therefore, that encoding 
provides an executable mechanism that integrates satisfactorily the dual semantic 
representation of communication acts.

In the following three subsections we present respectively the main features of  the 
\textsc{CommOnt} ontology, the dynamic semantics associated to communication acts, 
and an explanation of how these formalisms interact.

\subsection{Main features of the \textsc{CommOnt} ontology}

The goal of the \textsc{CommOnt} ontology is to facilitate the interoperation 
among agents belonging to
different Information Systems.
The leading categories of that ontology are: first,
\textit{communication acts} that are used for interaction by
\textit{actors} and that have different purposes and deal with
different kinds of contents; and second, \textit{contents}, which are
the sentences included in the communication acts.

The main design criteria adopted for the communication acts category
of the \textsc{CommOnt} ontology is to follow the \textit{speech
acts} theory \citep{Austin62}, a linguistic theory that is
recognized as the principal source of inspiration for designing the
most familiar standard agent communication languages.  Following that
theory, every communication act is the sender's expression of an
attitude toward some possibly complex proposition. A sender performs a
communication act, which is expressed by a coded message, and is
directed to a receiver. Therefore, a communication act has two main
components.  First, the attitude of the sender, which is called the
\textit{illocutionary force} (\textit{F}), that expresses social
interactions such as informing, requesting or promising. And second, 
the \textit{propositional content} (\textit{p}), 
which is the subject of the attitude.
In \textsc{CommOnt}  this \textit{F(p)} framework is followed, and 
different kinds of illocutionary forces and contents
leading to different classes of communication acts are described. 

\textsc{CommOnt} is divided into three interrelated layers: 
\textit{upper}, \textit{standards} and \textit{applications}, that group communication acts at
different levels of abstraction. 
\textsc{CommOnt} terminology is described using \textsc{OWL-DL}, the description logic 
profile of the Web
Ontology Language \textsc{OWL}\citep{OWL}. Therefore, communication acts among
agents that commit to \textsc{CommOnt} have an abstract
representation as individuals of classes which are specializations of a shared 
universal class of communication acts.

In the upper layer ---according to Austin's speech acts theory--- five upper classes of 
communication acts 
corresponding to \textit{Assertives}, \textit{Directives}, \textit{Commissives}, \textit{Expressives} 
and \textit{Declaratives} are specified. In addition, the top class \texttt{CommunicationAct}\footnote{\texttt{This type} 
style refers to terms specified in the ontology.} is defined, which represents 
the universal class of communication acts. Every particular communication act is 
an individual of this class. 
In \textsc{CommOnt}, components of a class are represented by properties. 
 The most important properties of \texttt{CommunicationAct} are the content 
and the actors who send and receive the communication act.
Following is DL axiom representing that structure can be found:

\scriptsize
\begin{eqnarray*}
\texttt{CommunicationAct} & \sqsubseteq & \texttt{=1 hasSender} \sqcap \forall \texttt{hasSender.Actor} \sqcap \forall \texttt{hasReceiver.Actor}\\
& & \sqcap \forall \texttt{hasContent.Content} 
\end {eqnarray*}
\normalsize

 There are some other properties related
to the context of a communication act such as the conversation in which it is inserted or 
a link to the domain ontology that includes the terms used in the content, but those 
details are out of the scope of this paper. In addition to the principal classes 
\texttt{Assertive}, \texttt{Directive}, \texttt{Commissive}, 
and \texttt{Declarative}, some other interesting subclasses are defined. For example: 

\scriptsize
\begin{eqnarray*}
\texttt{Request} & \sqsubseteq & \texttt{Directive} \sqcap \exists \texttt{hasContent.Command}\\
\texttt{Accept} & \sqsubseteq & \texttt{Declarative}\\
\texttt{Responsive} & \sqsubseteq & \texttt{Assertive} \sqcap \exists \texttt{inReplyTo.Request}\\
\texttt{Inquiry} & \sqsubseteq & \texttt{Directive} \sqcap \exists \texttt{hasContent.ReportAct}
\end {eqnarray*}
\normalsize

A standards layer extends the upper layer of the ontology with specific terms that represent
classes of communication acts of general purpose agent communication languages,
like those from \textsc{KQML} or \textsc{FIPA-ACL}.
Although the semantic framework of those agent communication languages
may differ from the semantic framework adopted in \textsc{CommOnt}, in our opinion 
enough basic concepts and principles are shared to such an extent that a commitment to
ontological relationships can be undertaken in the context of the interoperation of
Information Systems.

With respect to \textsc{FIPA-ACL}, we can observe that it proposes four 
primitive communicative acts \citep{FIPA-ACL}: \textit{Confirm}, \textit{Disconfirm},
\textit{Inform} and \textit{Request}. 
The terms \texttt{FIPA-Confirm}, \texttt{FIPA-Disconfirm}, \texttt{FIPA-Inform} and 
\texttt{FIPA-Request} are used to represent them as classes in \textsc{CommOnt}. Furthermore, 
the rest of the {\textsc FIPA} communicative acts are derived from these mentioned four 
primitives. Following are some examples of descriptions:

\scriptsize
\begin{eqnarray*}
\texttt{FIPA-Request} & \sqsubseteq & \texttt{Request}\\
\texttt{FIPA-Query-If} & \sqsubseteq & \texttt{Inquiry} \sqcap \texttt{FIPA-Request} \sqcap \texttt{=1 hasContent} \sqcap\\
& & \forall\texttt{hasContent.InformIf}\\
\texttt{FIPA-Agree} & \sqsubseteq & \texttt{Accept}
\end {eqnarray*}
\normalsize

Analogously, communication acts from {\textsc KQML} can be analyzed and the corresponding terms
can be specified in \textsc{CommOnt}. It is vital for the interoperability aim to have the ability to 
specify ontological relationships among classes of different standards. For example:

 \scriptsize
\begin{eqnarray*}
\texttt{KQML-Ask-If} & \equiv & \texttt{FIPA-Query-If} \\
\texttt{KQML-Tell} & \equiv & \texttt{FIPA-Inform}\\
\texttt{KQML-Achieve} & \equiv & \texttt{FIPA-Request} \sqcap \exists\texttt{hasContent.Achieve}
\end {eqnarray*}
\normalsize

Finally,
it is often the case that every Information System uses a limited
collection of communication acts that constitute its particular agent communication language. 
The applications layer reflects the terms describing communication acts used in such 
particular Information Systems.
The applications layer of the \textsc{CommOnt} ontology provides a framework for the 
description of the nuances of such communication acts. 
Some of these communication acts can be defined as specializations
of existing classes in the standards layer and some others as specializations of
upper layer classes. For example:

 \scriptsize
\begin{eqnarray*}
\texttt{A-MedicineModify} & \equiv & \texttt{Request} \sqcap \texttt{=1 hasContent} \sqcap \\
& & \forall\texttt{hasContent.(Overwrite } \sqcap \exists \texttt{hasSubject.Medicine)}\\
\texttt{A-MedicineModify} & \sqsubseteq & \texttt{FIPA-Request}\\
\texttt{A-VitalSignInform} & \sqsubseteq & \texttt{Responsive} \sqcap \texttt{=1 hasContent} \sqcap \\
& & \forall\texttt{hasContent.(Proposition } \sqcap \exists \texttt{hasSubject.VitalSignData)} \sqcap\\
& & \texttt{=1 inReplyTo} \sqcap \forall\texttt{inReplyTo.VitalSignQuery}\\
\texttt{A-VitalSignInform} & \sqsubseteq & \texttt{FIPA-Inform}
\end {eqnarray*}
\normalsize
Interoperation between agents of two systems using different kinds of communication acts 
will proceed through these upper and standard layer classes. 

In summary, \textsc{CommOnt} provides a terminology for communication acts and formal 
term relationships of equivalence and subsumption that allow to reason for interoperability 
purposes.


%
\subsection{Dynamic semantics associated to communication acts}\label{semantics}
Communication acts have received formal semantics based on mental concepts such as 
\textit{beliefs}, \textit{desires} and \textit{intentions}. However, that option has been 
criticized for the approach \citep{Singh98} as well as on its analytical difficulties 
\citep{Wooldridge00}. By contrast, we have adopted the so called social approach \citep{Singh00,Venkatraman99,Fornara02} 
to express the dynamic semantics of communication acts described in the \textsc{CommOnt} ontology. According to 
the social approach, when agents interact they become involved in social commitments or obligations to 
each other. Those commitments are public, and therefore they are 
suitable for an objective and verifiable semantics of agent interaction.

\begin{definition}
A \textit{base-level commitment} \textsf{C}(\textit{x, y, p}) is a ternary relation representing 
a commitment made by \textit{x} (the \textit{debtor}) to \textit{y} (the \textit{creditor}) to bring 
about a certain proposition \textit{p}.
\end{definition} 

Sometimes an agent accepts a commitment only if a certain condition holds or, interestingly, only 
when a certain commitment is made by another agent. This is called a conditional commitment.

\begin{definition}
A \textit{conditional commitment} \textsf{CC}(\textit{x, y, p, q}) is a quaternary relation 
representing that if the condition \textit{p} is brought out, \textit{x} will be committed to 
\textit{y} to bring about the proposition \textit{q}.
\end{definition}

Moreover, the formalism we use for reasoning about commitments is based on the Event Calculus, 
which is a logic-based 
formalism for representing actions and their effects. The basic ontology of the Event Calculus 
comprises \textit{events}, \textit{fluents} and \textit{time points}: \textit{events} 
correspond to \textit{actions} in our context; \textit{fluents} are predicates whose truth 
value may change over time. Event calculus  
includes predicates for saying what happens and when (\textit{Happens}), for describing the 
initial situation (\textit{Initially}), for describing 
the effects of actions (\textit{Initiates} and \textit{Terminates}), 
and for saying what fluents hold at what times (\textit{HoldsAt}). 
See~\citep{Shanahan99} for more explanations.

Commitments (base-level and conditional) can be considered fluents, and semantics of 
communication acts can be expressed with predicates in the Event Calculus. 

For example, following there are some predicates that describe the semantics associated to the 
classes of communication acts \texttt{Request}, \texttt{Accept} and \texttt{Responsive}, 
which appear in the upper level of \textsc{CommOnt} and whose descriptions have been shown 
in the previous subsection.
The semantics is determined by the fluents that are initiated or terminated as a result of  
delivering a message of that class from a sender to a receiver. The set of fluents that 
hold at a moment describe the state of the interaction. 

\begin{itemize}
\item 
\textit{Initiates(Request(s, r, P)}, \textsf{CC}\textit{(r, s, accept(r, s, P), P), t)}. \\ 
A Request from \textit{s} to \textit{r} produces the effect of generating a conditional 
commitment expressing that if the receiver \textit{r} accepts the demand, it will be commited 
to the proposition \textit{P} in the content of the communication act.
\item 
\textit{Initiates(Accept(s, r, P), accept(s, r, P), t)}.\\
The sending of an Accept produces the effect of generating the accept fluent.
\item
\textit{Terminates(Responsive(s, r, P, RA),}\textsf{C}\textit{(s, r, RA), t)}.\\
\textit{Terminates(Responsive(s, r, P, RA),}\textsf{CC}\textit{(s, r, accept(s, r, RA), RA), t)}.\\
By sending a message of the class Responsive, the commitment (either base-level or conditional) of the sender \textit{s} towards 
the receiver \textit{r} to bring about proposition \textit{RA} ceases to hold.
\end{itemize}
In summary, communication acts have a dual semantic representation. They are described in 
\textsc{CommOnt} in terms of their structure and hierarchical relationships and moreover there are 
some Event Calculus predicates which specify their effects.

Furthermore, some rules are needed to capture the dynamics of commitments.
Commitments 
are a type of fluent, typically put in force by communication acts, that become inoperative after the 
appearance of other fluents. In the following rules \textit{e(x)} represents an event caused by 
\textit{x}. The first rule declares that when a debtor of a commitment that is in force causes an event that 
initiates the committed proposition, the commitment ceases to hold.

\textsc{Rule 1:}  
\textit{HoldsAt(}\textsf{C}\textit{(x, y, p), t)} $\wedge$ \textit{Happens(e(x), t)} $\wedge$ \textit{Initiates(e(x), p, t)} $\rightarrow$ 
\textit{Terminates(e(x),}\textsf{C}\textit{(x, y, p), t)}.

The second rule declares that a conditional commitment that is in force disappears and generates a base-level 
commitment when the announced condition is brought out by the creditor.

\textsc{Rule 2:}  
\textit{HoldsAt(}\textsf{CC}\textit{(x, y, c, p), t)} $\wedge$ \textit{Happens(e(y), t)} $\wedge$ \textit{Initiates(e(y), c, t)} $\rightarrow$
\textit{Initiates(e(y),}\textsf{C}\textit{(x, y, p), t)} $\wedge$ 
\textit{Terminates(e(y),}\textsf{CC}\textit{(x, y, c, p), t)}.
%

\subsection{Encoding of the interaction of the dual semantic representation} 

SWRL is a combination of OWL DL with the Unary/Binary Datalog RuleML sublanguages of 
the Rule Markup Language. Therefore, \textsc{CommOnt} 
axioms can be managed naturally. Event calculus predicates and rules can be sistematically 
encoded in SWRL using a reification technique. For example, action 
\textit{Request(s, r, P)} is encoded as an individual \textit{x} in class \texttt{Request} 
and with \textit{s, r} and \textit{P} as values for its corresponding object properties: 
\texttt{hasSender}, \texttt{hasReceiver} and \texttt{hasContent} (specified in 
\textsc{CommOnt}). Analogously, a fluent \textit{accept(r, s, P)} may be represented  with 
the following assertions about individuals: 
\textit{Acceptance(a)}, \textit{hasSignatory(a,r)}, \textit{hasAddressee(a,s)}, and 
\textit{hasObject(a,P)}.

Then, the aforementioned rules, as well as effects of the application of communication acts, 
can be encoded with SWRL rules. 
For instance, the predicate 
\textit{Initiates(Request(s, r, P), }\textsf{CC}\textit{(r, s, accept(r, s, P), P), t)} 
can be encoded as follows\footnote{We are aware that in the human readable syntax of SWRL, variables are prefixed with a question mark (e.g. \textit{?x}). However, for the sake of visual clarity, the question mark has been removed from all the SWRL variables in this paper}:

\textit{Request(x)} $\wedge$ \textit{hasSender(x,s)} $\wedge$ \textit{hasReceiver(x,r)} $\wedge$ \textit{hasContent(x,p)} $\wedge$ \textit{hasCommit(x,c)} $\wedge$ \textit{isConditionedTo(c,a)} $\rightarrow$ \textit{initiates(x,c)} $\wedge$ \textit{hasDebtor(c,r)} $\wedge$ \textit{hasCreditor(c,s)} $\wedge$ \textit{hascondition(c,p)} $\wedge$ \textit{Acceptance(a)} $\wedge$ \textit{hasSignatory(a,r)} $\wedge$ \textit{hasAddressee(a,s)} $\wedge$ \textit{hasObject(a,p)}

\section{Representations of the semantics of protocols}\label{description} 
Similarly to the representation of communication acts, presented in the previous section, 
we propose a dual representation for protocols. On the one hand, we define a 
structure-based representation, using OWL-DL descriptions; on the other hand, 
we define a fluent-based semantics of protocols.

\subsection{OWL-DL description of protocols}
As mentioned in the previous section, we use STS as models for representing protocols. 
More specifically, in this paper 
we restrict our work to deterministic STS without cycles.
In order to represent those models using OWL-DL, we have defined 
five different classes: \texttt{Protocol}, \texttt{State}, \texttt{Transition}, 
\texttt{Fluent} and \texttt{Commitment}, which respectively represent protocols, states, transitions in protocols, and fluents and commitments associated to states. 
With these five classes and some (specialized) subclasses we are able to represent 
enough structure in order to fully describe the components of individual instances 
of our protocols. 

We model those class descriptions with the following guidelines. 
Fluents are associated to states where they hold\footnote{$\exists$\texttt{hasFluent}$\sqsubseteq$ \texttt{State} and $\exists$\texttt{hasFluent}$^{-} \sqsubseteq$ \texttt{Fluent} mean that the class \texttt{State} and the class \texttt{Fluent} are respectively the domain and range of the property \texttt{hasFluent}.}: 

\scriptsize
\begin{eqnarray*}
\exists \texttt{hasFluent} & \sqsubseteq & \texttt{State} \ \ ;\ \ \exists \texttt{hasFluent$^{-}$} \sqsubseteq \texttt{Fluent}
\end{eqnarray*}
\normalsize
Transitions get out of states and every transition is labelled with the communication 
act that is delivered by that event, and it is associated to the state reached by 
that transition:

\scriptsize
\begin{eqnarray*}
\exists \texttt{hasTransition} & \sqsubseteq & \texttt{State} \ \ ;\ \ \exists \texttt{hasTransition$^{-}$} \sqsubseteq \texttt{Transition}\\
  \texttt{Transition}& \equiv & \texttt{=1 hasCommAct} \sqcap \texttt{=1 hasNextState} \\
\exists \texttt{hasCommAct} & \sqsubseteq & \texttt{Transition} \ \ ;\ \ \exists \texttt{hasCommAct$^{-}$} \sqsubseteq \texttt{CommunicationAct}\\
\exists \texttt{hasNextState} & \sqsubseteq & \texttt{Transition} \ \ ;\ \ \exists \texttt{hasNextState$^{-}$} \sqsubseteq \texttt{State}
\end{eqnarray*}
\normalsize
A protocol is an individual of the class \texttt{Protocol} and it is determined by 
the properties of its initial state, due to our conceptual modeling of 
states and transitions.

\scriptsize
\begin{eqnarray*}
\texttt{Protocol}& \equiv & \exists \texttt{hasInitialState.State} \\
\exists \texttt{hasInitialState} & \sqsubseteq & \texttt{Protocol} \ \ ;\ \ \exists \texttt{hasInitialState$^{-}$} \sqsubseteq \texttt{State}\\
\end{eqnarray*}
\normalsize
Some other interesting subclasses are specified in order to describe the elements 
that compose our protocols:

\scriptsize
\begin{eqnarray*}
\texttt{FinalState}& \sqsubseteq & \texttt{State} \\
 \texttt{Commitment}&\sqsubseteq& \texttt{Fluent} \sqcap  \texttt{ =1 hasDebtor} \sqcap \forall\texttt{hasDebtor.Actor} \sqcap \texttt{=1 hasCreditor} \sqcap\\
& & \forall\texttt{hasCreditor.Actor} \sqcap \texttt{=1 hasCondition} \sqcap \forall\texttt{hasCondition.Fluent}\\
  \texttt{ConditionalCommitment}&\sqsubseteq& \texttt{Fluent} \sqcap  \texttt{ =1 hasDebtor} \sqcap \forall\texttt{hasDebtor.Actor} \sqcap \texttt{=1 hasCreditor} \sqcap\\
& & \forall\texttt{hasCreditor.Actor} \sqcap \texttt{=1 hasCondition} \sqcap \forall\texttt{hasCondition.Fluent} \sqcap\\
 & &  \texttt{=1 isConditionedTo} \sqcap \forall\texttt{isConditionedTo.Fluent}
\end{eqnarray*}
\normalsize

We are conscious that alternative descriptions may be considered, but our conceptual 
modeling is disposed to favor the rule encoding of dynamic aspects of a protocol, 
as will be shown in the next subsection.

\subsection{Fluent-based semantics of protocols}\label{semantics}
 One of the most common approaches for comparing two protocols involves the discovery of structural relationships between them. However, we believe that dealing only with structural relationships is too rigid if a flexible interoperation among agents that use different standard protocols is to be promoted.
 For that reason, we propose to obtain an additional description of protocols, represented by their final states and the fluents that hold at those final states.
In order to do so, we propose to consider the following definitions.

Let $\mathcal{WR}$ be the set of SWRL rules encoding \textsc{Rule 1} and \textsc{Rule 2} 
presented in the previous section. Let $G$ and $G'$ be sets of fluents; and let $(s,l,s')$ be a 
transition in a STS. Then, $G:\textless(s,l,s')\textgreater\vdash_{\mathcal{WR}}G'$ is a 
\textit{transition derivation}, which means that, in the 
context of $\mathcal{WR}$, if $G$ is the set of fluents that hold at state $s$, when transition 
$(s,l,s')$ happens, $G'$ is the set of fluents that hold at state $s'$.

Transition derivations represent the dynamics of a protocol. When a transition is 
accomplished some fluents may become true, others may become false, and the rest 
remain unchanged.
 
The encoding of $G:\textless(s,l,s')\textgreater\vdash_{\mathcal{WR}}G'$ into SWRL rules is done by taking into account our OWL-DL descriptions of STS presented in the previous subsection. Two main rules have been defined: On the one hand, the fluent attachment rule attaches to state $s'$ the fluents initiated as a result of sending the communication act $l$:

\textit{Transition(t)} $\wedge$ \textit{hasNextState(t,s')} $\wedge$ \textit{hasCommAct(t,l)} $\wedge$ \textit{initiates(l,f)} $\rightarrow$ \textit{hasFluent(s',f)}

On the other hand, the fluent transmission rule transfers the fluents that hold in state $s$ and that must also hold in state $s'$ because the act of sending the communication act $l$ has no effect on them:

\textit{hasFluent(s, f)} $\wedge$ \textit{hasTransition(s, t)} $\wedge$ \textit{hasNextState(t, s')} $\rightarrow$ \textit{hasFluent(s', ?f)}

\begin{definition}
A $branch$ of a protocol $P=(S,s_{0},L,T,F)$ is a sequence of transitions from $T$, 
\textless($s_{i-1}, l_{i}, s_{i}$)$>_{i=1..n}$, that begins in the initial state $s_{0}$ 
and ends in a final state $s_{n}\in$ $F$. 
We denote $\Omega(P)$ to the set of branches of protocol $P$.
\end{definition}


Let $B=\textless(s_{i-1}, l_{i}, s_{i})\textgreater_{i=1..n}$ be a branch, then $G_{0}:B\vdash_{\mathcal{WR}}G_{n}$ means that there exists the sets of fluents $G_{i}$ $i=1..n$ such that $G_{i-1}:\textless(s_{i-1}, l_{i}, s_{i})\textgreater\vdash_{\mathcal{WR}}G_{i}$

\begin{definition}
If $B$ is a branch of protocol $P$($S$, $s_{0}$, $L$, $T$, $F$), $G_{0}$ is the set of fluents that hold in $s_{0}$ and $G_{0}:B\vdash_{\mathcal{WR}}G_{n}$, then $G_{n}$ is a \textit{protocol trace}. We denote $\mathcal{T}(B)$ to the final set of fluents generated by $B$. That is to say, $\mathcal{T}(B)=G_{n}$. 
\end{definition}

Notice that protocol traces are defined in terms of the semantics of communication acts, 
taking into account the content of the communication and not only the type of communication. 
By contrast, many other related works (see section \ref{trabajos}) consider only 
communication acts as atomic acts without considering their content or their semantics. 

From our viewpoint, the semantics of a protocol is determined by the traces of the protocol. 
That is to say, from a set-theoretical approach  
$\{\mathcal{T}(B)|\ B\in\Omega(P)\}$ is an interpretation of protocol $P$.

\section{Algorithm}\label{Algorithm} 

In this section we present the main steps of the algorithm that compares two protocols. As the semantics of the protocol is determined by ist traces, in order to perform the comparison we need to deal with the braches of the protocol. The algorithm comprises three main steps:
\begin{enumerate}
	 \item Separation of each protocol into branches: each protocol is separated into all the branches that can be generated from the initial state to a final state.
	 \item Generation of protocol traces: the protocol traces corresponding to each branch of each protocol are calculated.
	\item Pairing up branches: the branches of one of the protocols are compared to the branches of the other protocol.
\end{enumerate}

Moreover, once the previous steps have been concluded, several relationships between both protocols can be established from a global point of view.

Now we will develop these steps more thoroughly.

\subsection{Separation of the protocol into branches}\label{Separation}

In the first step, the protocol is separated into all the branches that can be generated from the initial state to a final state.

\begin{figure}
\begin{center}	
	\includegraphics[width=2.3in]{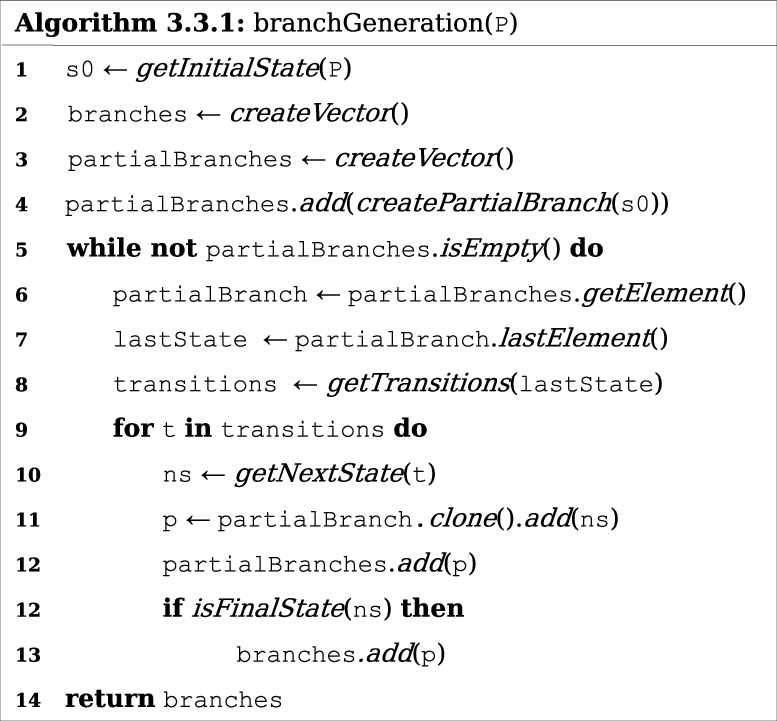}
	\caption{\label{algorithm2} Algorithm: Generation of branches}
\end{center}
\end{figure}

\begin{figure}
	\centering
	\includegraphics[width=3.5in]{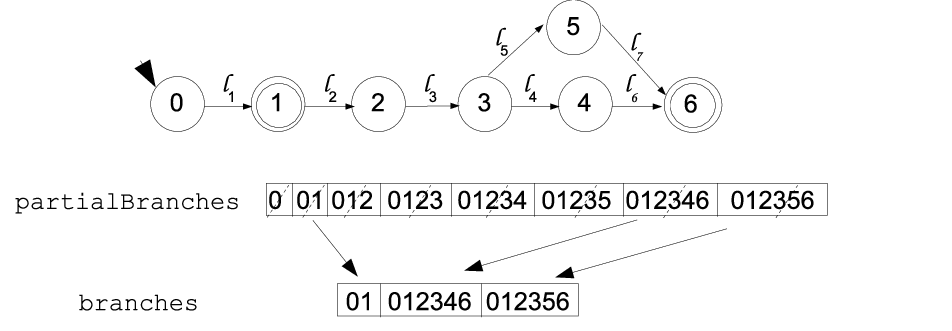}
	\caption{Separation of branches}
	\label{branchGeneration}
\end{figure}

 Algorithm 3.3.1 (see Fig.~\ref{algorithm2}) shows how to do that. In line 2, a global vector (\texttt{branches}) is created, which will be used to store the different branches of a protocol P after they have been calculated. In line 3 another vector (\texttt{partialBranches}) is created. This vector will be used as an auxiliary to store the partial branches that are generated in the process of calculating the complete branches. This vector is initialized, in line 4, with a partial branch that contains only the initial state of the protocol (s0). Taking into account the transitions of protocol P, this partial branch is repeatedly modified by adding new states to it (lines 5 to 13). Moreover, once one of the partial branches is completed (i.e. a final state has been reached), it is added to the global vector \texttt{branches} (lines 12-13), which will be returned once all the partial branches have been completed. An example can be found in Fig.~\ref{branchGeneration}.

\subsection{Generation of protocol traces}\label{Generation}

Given a branch $B$, with a simulation of its corresponding $G_{0}:B\vdash_{\mathcal{WR}}G_{n}$, the corresponding trace $\mathcal{T}(B)$ is calculated.
During that simulation the SWRL rules that encode the semantics of the communication acts (see section~\ref{semantics}) appearing in the branch are applied. Then, the set of fluents that hold at the final state of each branch is the corresponding trace. That set represents the effects of the protocol branch. 

\begin{figure}
	\centering
	\includegraphics[width=2.5in]{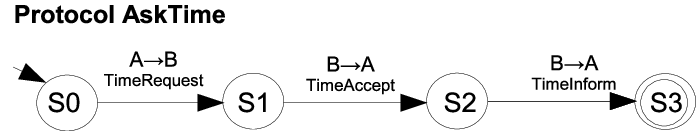}
	\caption{Protocol diagram} \label{aplicarReglas}
\end{figure}

For example, let us take protocol AskTime in Fig.~\ref{aplicarReglas}.
By using this protocol, agent A intends to get from agent B information about the time. So, in the first transition, agent A requests about the time. Then, agent B sends a message accepting to give the information about the time, and finally, in the third transition, agent B sends the requested information. The descriptions in the \textsc{CommOnt} ontology of the communication acts that appear in the protocol AskTime are the following:

\scriptsize
\begin{eqnarray*}
 \texttt{TimeRequest} & \equiv & \texttt{Request} \sqcap \texttt{=1 hasContent} \sqcap \forall\texttt{hasContent.TimeReq}\\
 \texttt{TimeAccept} & \equiv & \texttt{Accept} \sqcap \texttt{=1 hasContent} \sqcap \forall\texttt{hasContent.TimeReq}\\
 \texttt{TimeInform} & \equiv & \texttt{Responsive} \sqcap \texttt{=1 hasContent} \sqcap \forall\texttt{hasContent.TimeInfo} \sqcap \texttt{=1 inReplyTo} \sqcap\\
& & \forall\texttt{inReplyTo.TimeRequest}\\
\end{eqnarray*}
\normalsize

\begin{figure}
	\centering
	\includegraphics[width=2.5in]{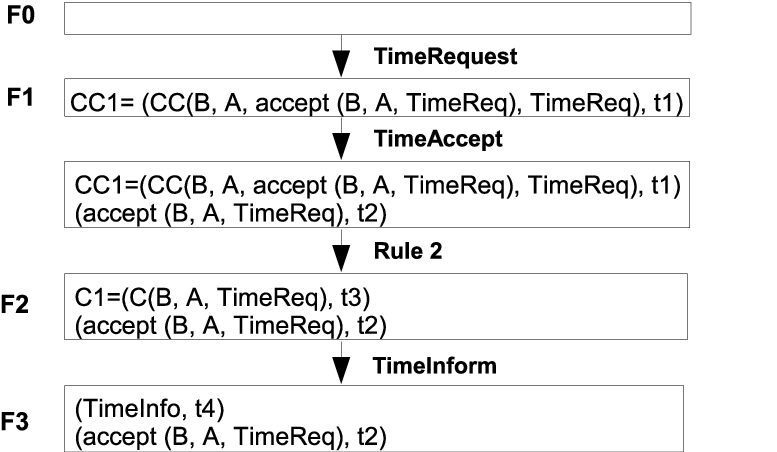}
	\caption{Protocol fluents} \label{fluentes}
\end{figure}

In Fig.~\ref{fluentes} we show which fluents are associated to the states of the protocol and how they vary as a consequence of the communication acts that are sent and the rules described in section~\ref{semantics}.
\normalsize
We depart from a situation where the set of fluents is empty (F0). The impact the communication acts of the protocol have over the set of fluents that hold at each moment is explained next:

\medskip
\texttt{\textbf{\underline{TimeRequest: A$\rightarrow$B}}}
\medskip

When the \texttt{TimeRequest} message is sent, the conditional commitment \texttt{CC1} 
is initiated (F1), which states that if agent B accepts to give information about the time, then it will be committed to do so. This happens because according to the definitions in the \textsc{CommOnt} ontology, \texttt{TimeRequest} is a subclass of \texttt{Request}, and as a result, the absatract predicate \textit{Initiates(Request(s, r, P), \textsf{CC}(\textit{r, s, accept(r,s,P), P}), t)} can be applied. More precisely, in this case the predicate is instanciated as  \textit{Initiates(Request(A, B, TimeReq), \textsf{CC}(\textit{B, A, accept(B,A,TimeReq), TimeReq}), $t_{1}$)}. The following SWRL rule, which encodes the abstract predicate, can be executed at this moment:

\textit{Request(x)} $\wedge$ \textit{hasSender(x,s)} $\wedge$ \textit{hasReceiver(x,r)} $\wedge$ \textit{hasContent(x,p)} $\wedge$ \textit{hasCommit(x,c)} $\wedge$ \textit{isConditionedTo(c,a)} $\rightarrow$ \textit{initiates(x,c)} $\wedge$ \textit{hasDebtor(c,r)} $\wedge$ \textit{hasCreditor(c,s)} $\wedge$ \textit{hascondition(c,p)} $\wedge$ \textit{Acceptance(a)} $\wedge$ \textit{hasSignatory(a,r)} $\wedge$ \textit{hasAddressee(a,s)} $\wedge$ \textit{hasObject(a,p)}

This rule initiates the aforementioned conditional commitment \texttt{CC1}, and due to the fluent attatchment rule (section \ref{semantics}, \texttt{CC1} is attached to state S1, and thus it holds in F1.

 \medskip
\texttt{\textbf{\underline{TimeAccept: B$\rightarrow$A}}}
\medskip

 Then, agent B agrees to respond by sending the \texttt{TimeAccept} message. Due to the predicate \textit{Initiates(Accept(s,r,P), accept(s,r,P), t)} --- instanciated in this case as \textit{Initiates(Accept(B,A,TimeReq), accept(B,A,TimeReq), $t_{2}$)}--- and the fact that \texttt{TimeAccept} is a subclass of \texttt{Accept}, the rule

\textit{Accept(x)} $\wedge$ \textit{hasSender(x,s)} $\wedge$ \textit{hasReceiver(x,r)} $\wedge$ \textit{hasContent(x,p)} $\wedge$ \textit{hasCommit(x,c)} $\rightarrow$ \textit{initiates(x,c)} $\wedge$ \textit{Acceptance(c)} $\wedge$ \textit{hasSignatory(c,s)} $\wedge$ \textit{hasAddressee(c,r)} $\wedge$ \textit{hasObject(a,p)}

is executed and consequently, the fluent \textit{accept(B, A, TimeReq)} is initiated. 
At this point, \textsc{Rule 2} (see section~\ref{semantics}) can be applied because \textit{accept(B,A,TimeReq)} is the condition of the conditional commitment \texttt{CC1}. As a consequence, \texttt{CC1} is terminated and the base commitment \texttt{C1} is initiated and attached to state S2 due to the aforementioned fluent attachment rule. We would like to remark that, since SWRL does not offer any primitive for retraction, the deletion of the fluent \texttt{CC1} is performed using the methods of the OWL-API\citep{Horridge07} that allow the manipulation of the ABox.

\medskip
\texttt{\textbf{\underline{TimeInform: B$\rightarrow$A}}}
\medskip

Finally, agent B sends the \texttt{TimeInform} message. As \texttt{TimeInform} is a subclass of \texttt{Responsive} and because of the predicate \textit{Initiates(Responsive(s,r,P, RA), P, t)}, ---instanciated as \textit{Initiates(Responsive(B,A,TimeInfo, TimeReq), TimeInfo, $t_{4}$)}--- the fluent \textit{TimeInfo}, is initiated. Moreover, \texttt{C1} is terminated due to the predicate \textit{Terminates(Responsive(s,r,P, RA), \textsf{C}(\textit{s,r,RA}), t)}, instanciated as \textit{Terminates(Responsive(B, A, TimeInfo, TimeReq), \textsf{C}(\textit{B,A,TimeReq}), $t_{4}$)}. In our environment, these predicates have been encoded with the rule:

\textit{Responsive(x)} $\wedge$ \textit{hasContent(x,p)} $\wedge$ \textit{inReplyTo(x,irt)} $\wedge$ \textit{initiates(irt, f)} $\rightarrow$ \textit{terminates(x,f)} $\wedge$ \textit{initiates(x, p)}

In addition, due to the fluent transmission rule, the fluent \textit{accept(B, A, TimeReq)} is transferred from state S2 to state S3.
So, at this point (F3) we can say that the fluents that hold at the final state of the protocol 
are (\textit{accept(B, A, TimeReq)}, $t_{2}$)  and (\textit{TimeInfo}, $t_{4}$) and so, the protocol trace 
[(\textit{accept(B, A, TimeReq)}, $t_{2}$), (\textit{TimeInfo}, $t_{4}$)] 
is \textit{generated} for the branch. 
As our algorithm does not make use of the \textit{time} parameter, it will be omitted in the following analysis.

\subsection{Pairing up branches from two different protocols} \label{Pairing}
Once protocol traces have been generated, it is necessary to establish 
relationships between branches of the two protocols $P1$ and $P2$ being compared. 
In order to 
compare two branches, we compare their protocol traces, and thus, their
 fluents. To do so, the algorithm first evaluates the cartesian product of 
branches $\Omega(P1)\times \Omega(P2)$, taking into account the following considerations:

Given $B1\in\Omega(P1)$ and $B2\in\Omega(P2)$,  
four separate cases may occur when comparing two fluents $t1$ $\in$ $\mathcal{T}(B1)$ and $t2$ $\in$ $\mathcal{T}(B2)$ :

\begin{description}
\item [(\textit{eq})] $t1$ and $t2$ are \textit{equivalent}: \textit{msc($t1$)} $\equiv$ \textit{msc($t2$)}, being \textit{msc(t)} the most specific concept of a fluent $t$ in regard to an ontology of fluents. In this case we could see those fluents as \textit{clones}, their names being their sole difference.\\
\item [(\textit{g1})] \textit{$t1$} is \textit{more general} than \textit{$t2$}: \textit{msc($t2$)} $\sqsubseteq$ \textit{msc($t1$)} and \textit{msc($t2$)} 
$\not\equiv$ \textit{msc($t1$)}.\\ 
\item [(\textit{g2})] \textit{$t2$} is \textit{more general} than \textit{$t1$}: \textit{msc($t1$)} $\sqsubseteq$ \textit{msc($t2$)} and \textit{msc($t1$)} $\not\equiv$ \textit{msc($t2$)}.\\
\item [(\textit{in})] \textit{$t1$} and \textit{$t2$} are \textit{incomparable}: 
\textit{msc($t1$)} $\not\sqsubseteq$ \textit{msc($t2$)} and 
\textit{msc($t2$)} $\not\sqsubseteq$ \textit{msc($t1$)}
(They have no relationship to each other).
\end{description}

It may happen that a fluent exists in $\mathcal{T}(B1)$ which is \textit{incomparable} 
with any fluent in $\mathcal{T}(B2)$, and a fluent in $\mathcal{T}(B2)$ which is 
\textit{incomparable} with any fluent in $\mathcal{T}(B1)$; then, we say that the 
pairing up of $B1$ and $B2$ is not \textit{feasible}.
 This means that $B1$ generates some fluent that is not generated by $B2$ and viceversa.
 
\begin{definition}
$(B1, B2) \in \Omega(P1)\times \Omega(P2)$ is a \textit{feasible pair} iff 
[$\forall t1 \in B1.\ \exists t2\in B2.\ t1$ and $t2$ satisfy $(eq), (g1)$ or $(g2)$] $\vee$ 
[$\forall t2 \in B2.\ \exists t1\in B1.\ t1$ and $t2$ satisfy $(eq), (g1)$ or $(g2)$]
\end{definition}

We are not interested in unfeasible pairs of branches because, as will be shown in 
section~\ref{ProtocolRelationships}, the relationships we are interested in are those where all 
the fluents in at least one of the traces must be related to some fluent in the other. 

Each feasible pair $(Bi, Bj) \in \Omega(P1)\times \Omega(P2)$ receives a 
4-place tuple \textit{valuation} ($x_{0}^{ij}$,$x_{1}^{ij}$,$x_{2}^{ij}$,$x_{3}^{ij}$) where:
\begin{itemize}
\item 
$x_{0}^{ij}$ is the number of pairs of fluents that fulfill case \textit{(eq)}.
\item 
$x_{1}^{ij}$ is the number of pairs of fluents that fulfill case \textit{(g1)}.
\item 
$x_{2}^{ij}$ is the number of pairs of fluents that fulfill case \textit{(g2)}.
\item 
$x_{3}^{ij}$ is the exceeding number of fluents: 
$x_{3}^{ij}$=$|\#\mathcal{T}(B1)-\#\mathcal{T}(B2)|$, 
where $\#\mathcal{T}(B1)$ and $\#\mathcal{T}(B2)$ are the number of fluents of $\mathcal{T}(B1)$ and $\mathcal{T}(B2)$ respectively.
\end{itemize}

Then, a \textit{similarity metric} is defined by means of the following function, 
which yields a value in $[0, 1]$ for every 4-place tuple:

$$
f(x_{0},x_{1},x_{2},x_{3}) = \frac{x_{0}+max(x_{1},x_{2})}{\sum_{i=0}^{3}x_{i}} 
$$

The intuition behind this formula is that a greater value represents more similarity. 
Our interest is to relate as many fluents as 
possible within a pair of branches, regardless of whether their relationship is of equivalence 
or subsumption (that is why $x_{0}$ is given the same weight as $max(x_{1}, x_{2})$ 
in the formula). Division by $\sum_{i=0}^{3}x_{i}$ assures the function value doesn't go 
beyond $1$.

\begin{definition}
A \textit{matching} $\pi$ is a subset of feasible pairs from $\Omega(P1)\times \Omega(P2)$ 
such that every branch from 
the smallest set of $\Omega(P1), \Omega(P2)$ is paired up, and any branch is paired up with at 
most another branch. ($\pi$ is the graph of an injective map.)
\end{definition}

Our aim is to obtain the matching that maximizes the sum of the similarity metric applied 
to the valuation of its pairs. Formally, we look for the matching $\pi$ that
\begin{displaymath}
\mathrm{maximizes} \sum_{(B_{i}, B_{j})\in \pi}f(x_{0}^{ij}, x_{1}^{ij}, x_{2}^{ij}, x_{3}^{ij})
\end{displaymath}

In the case that there exist two or more matchings $\pi_{k} (k\in 1\ldots n)$ that maximize 
the sum above, we priorize the equivalence relationships between fluents over subsumption 
relationships. Then, the best matching is anyone of the set $\{ \pi_{k}: (k\in 1\ldots n)\}$ 
such that 
\begin{displaymath}
\mathrm{maximizes} \sum_{(B_{i}, B_{j})\in \pi_{k}}g(x_{0}^{ij}, x_{1}^{ij}, x_{2}^{ij}, x_{3}^{ij}),\ \ \mathrm{with}\ g(x_{0},x_{1},x_{2},x_{3}) = \frac{x_{0}}{\sum_{i=0}^{3}x_{i}}
\end{displaymath}

\begin{figure}
	\centering
	\includegraphics[width=5.3in]{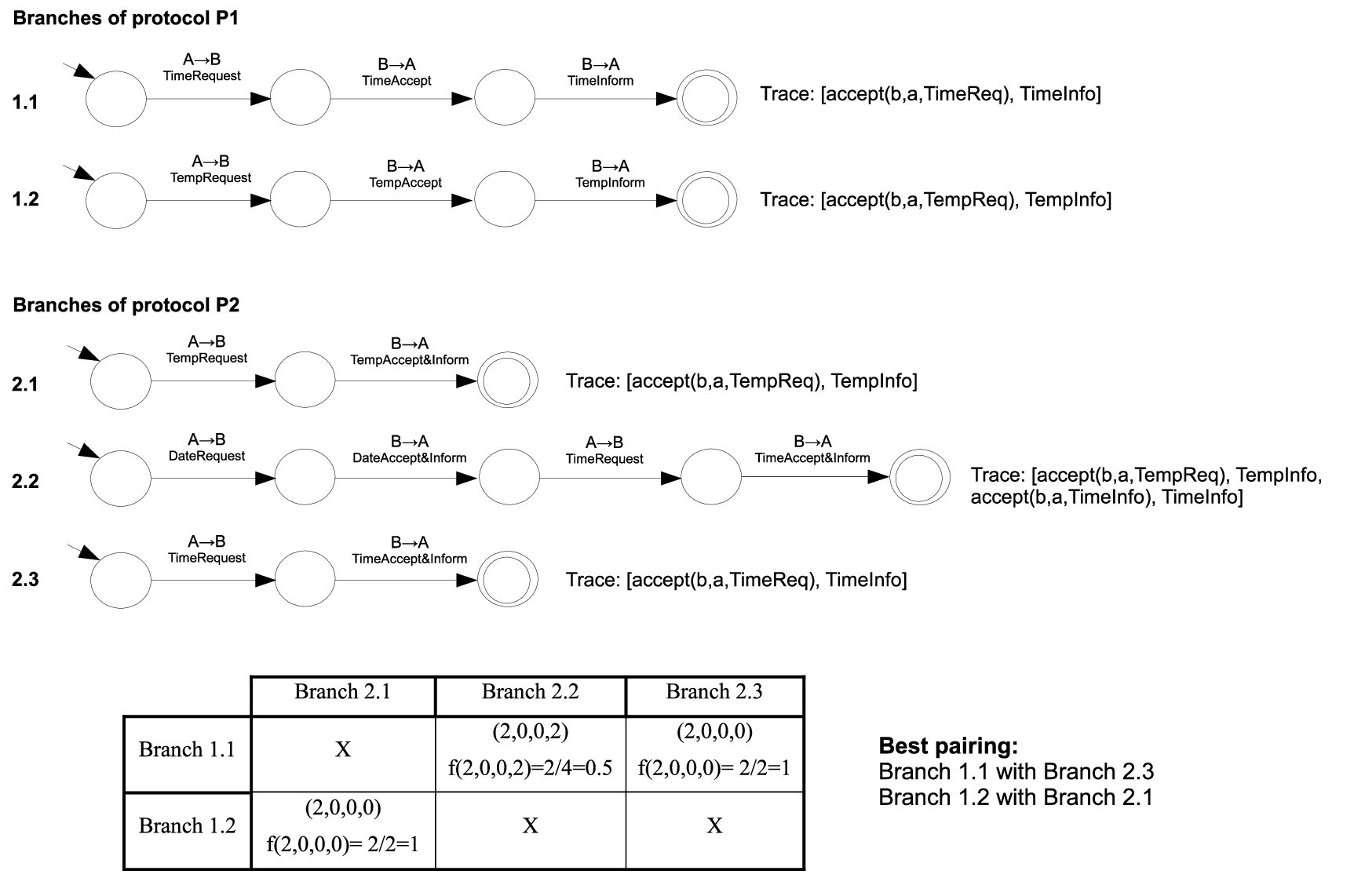}	
	\caption{\label{pairingExample}Pairing up branches of different protocols}
\end{figure}


In Fig.~\ref{pairingExample} we present an example of the process of pairing up branches. On the one hand, in protocol $P1$ agent A requests and later receives information about the time (branch 1.1) or about the temperature (branch 1.2). Three transitions are necessary in each of the branches to reach the final state of the protocol. On the other hand protocol $P2$ is composed of three branches: Branch 2.1 has the same semantics as branch 1.2 but in this case only two transitions are needed, since with the \texttt{TempAccept\&Inform} communication act, agent B both accepts and responds to the request. In branch 2.2, agent A requests for information about the date and the time and gets the respective responses (two transitions for each of the parameters). Finally, information about the time is requested in branch 2.3, also in two steps. The table in Fig.~\ref{pairingExample} registers the valuation and similarity metric value of the feasible pairs from $\Omega(P1)\times \Omega(P2)$. When a pair is not feasible we represent it with an \texttt{X} in the corresponding cell.

\section{Protocol relationships}\label{ProtocolRelationships}
Once the algorithm described in the previous section has finished the analysis of branches of protocols, the next step consists on discovering the semantic relationships between those protocols. This section presents the collection of semantic relationships we are considering. Firstly, we present the definitions of relationships between two protocol branches, followed by the definitions of relationships between protocols.

\subsection{Relationships between branches}
Let $A$, $B$ be two protocol branches, and $\mathcal{T}(A)$, $\mathcal{T}(B)$ be the protocol traces generated by branches $A$ and $B$ respectively.

\begin{definition}
Branch \textit{A} is a \textit{equivalent} to branch \textit{B} (\textit{A}=$_{b}$\textit{B}) if there exists a bijective function $\phi:\mathcal{T}(A)\rightarrow\mathcal{T}(B)$ such that 
$\forall t \in \mathcal{T}(A).\ msc(t)\equiv msc(\phi(t))$.
\end{definition}

\begin{definition}
Branch \textit{A} is a \textit{specialization} of branch \textit{B} (\textit{A}$\ll_{b}$\textit{B}) if there exists a bijective function $\phi:\mathcal{T}(A)\rightarrow\mathcal{T}(B)$ such that 
$\forall t \in \mathcal{T}(A).\ msc(t)\sqsubseteq msc(\phi(t))$.
\end{definition}

We denote \textit{Prune}[$B/s_{k}$] to the branch 
that results from pruning branch $B$ from state $s_{k}$ onwards. Given a branch 
$B=<(s_{i-1}, l_{i}, s_{i})>_{i=1..n}$, and $1\leq k\leq n$, \textit{Prune}[$B/s_{k}$]=
$<(s_{i-1}, l_{i}, s_{i})>_{i=1..k}$.

\begin{definition}
Branch \textit{A} is a \textit{prefix} of branch \textit{B} 
(\textit{A}$\mathcal{=}_{pre}$\textit{B}) if there exists a state $s_{k}$ in  \textit{B} such that \textit{A}$\mathcal{=}_{b}$\textit{Prune}[B/$s_{k}$].

Branch \textit{A} is a \textit{specialized-prefix} of branch \textit{B} 
(\textit{A}$\ll_{pre}$\textit{B}) if there exists a state $s_{k}$ in \textit{B} such that \textit{A}$\ll_{b}$\textit{Prune}[B/$s_{k}$].
\end{definition}

Let $s_k$($1\leq k<n$) a state in the definition of branch \textit{B}. We denote \textit{ChangeInit}[$B/s_{k}$] to the new branch that results from modifying the definition of branch \textit{B} in such a way that $s_{k}$ becomes the initial state of the new branch: $\textit{ChangeInit}[B/s_{k}]$=\textless($s_{i-1}$, $l_{i}$, $s_{i}$)\textgreater$_{i=k+1..n}$
 
\begin{definition}
Branch \textit{A} is a \textit{suffix} of branch \textit{B} (\textit{A}$\mathcal{=}_{suf}$\textit{B}) if there exists a state $s_{k}$ in \textit{B} such that \textit{A}$\mathcal{=}_{b}$\textit{ChangeInit}[B/$s_{k}$].

Branch \textit{A} is a \textit{specialized-suffix} of branch \textit{B} (\textit{A}$\ll_{suf}$\textit{B}) if there exists a state $s_{k}$ in \textit{B} such that \textit{A}$\ll_{b}$\textit{ChangeInit}[B/$s_{k}$].
\end{definition}

\begin{definition}
Branch \textit{A} is an \textit{infix} of branch \textit{B} (\textit{A}$\mathcal{=}_{inf}$\textit{B}) if there exists a state $s_{k}$ in \textit{B} such that \textit{A}$\mathcal{=}_{pre}$\textit{ChangeInit}[B/$s_{k}$] (i.e. \textit{A} is a prefix of a suffix of \textit{B}).

Branch \textit{A} is an \textit{specialized-infix} of branch \textit{B} (\textit{A}$\ll_{inf}$\textit{B}) if there exists a state $s_{k}$ in \textit{B} such that \textit{A}$\ll_{pre}$\textit{ChangeInit}[B/$s_{k}$]. 
\end{definition}

\begin{definition}
Branch \textit{A} is an \textit{complement\_to\_infix} of branch 
\textit{B} (\textit{A}$\mathcal{=}_{c}$\textit{B}) if there exists a state
 $s_{k}$ in \textit{A} such that \textit{Prune}[A/$s_{k}$]$\mathcal{=}_{pre}$\textit{B}
 and \textit{ChangeInit}[A/$s_{k}$]=$_{suf}$\textit{B}.

Branch \textit{A} is an \textit{specialized-complement\_to\_infix} of branch 
\textit{B} (\textit{A}$\ll_{c}$\textit{B}) if there exists a state $s_{k}$ in  \textit{A} such that \textit{Prune}[A/$s_{k}$]$\ll_{pre}$\textit{B} and \textit{ChangeInit}[A/$s_{k}$]$\ll_{suf}$\textit{B}.
\end{definition}

Taking into account the previous definitions we have developed a reasoning service
that decides if a branch is a prefix, a suffix, an infix or a complement\_to\_infix of another branch, either in an equivalent or specialization sense.

\subsection{Relationships between protocols}\label{RelationshipsBetweenProts}
Protocols are constituted by branches, and therefore we use the definitions in the previous subsection to define the relationships between protocols.

\begin{definition}
Protocol $P$ is \textit{equivalent} to protocol $Q$ ($P$[$\mathcal{E}$]$Q$) if there exists a bijective function $\phi:\Omega(P)\rightarrow\Omega(Q)$ such that 
$\forall A \in \Omega(P).\ A\mathcal{=}_{b}\phi(A)$.
\end{definition}

\begin{definition}
Protocol $P$ is a \textit{specialization} of protocol $Q$ ($P$[$\mathcal{Z}$]$Q$) if there exists a bijective function $\phi:\Omega(P)\rightarrow\Omega(Q)$ such that 
$\forall A \in \Omega(P).\ A\ll_{b}\phi(A)$.
\end{definition}

Sometimes, a protocol is defined by restrictions on the allowable communication acts at 
some states of a general protocol. In those situations the application of those restrictions is reflected in the corresponding effects.
\begin{definition}
Let $\Omega'(Q)$ a proper subset of $\Omega(Q)$ ($\Omega'(Q)\subset\Omega(Q)$). Protocol $P$ is a \textit{restriction} 
of protocol $Q$ ($P$[$\mathcal{R}$]$Q$) if there exists a bijective function 
$\phi:\Omega(P)\rightarrow\Omega'(Q)$ 
such that 
$\forall A \in \Omega(P).\ A\mathcal{=}_{b}\phi(A)$.
\end{definition}

\begin{definition}
Protocol $P$ is a \textit{prefix} of protocol $Q$ ($P$[$\mathcal{P}$]$Q$) if there exists a bijective function $\phi:\Omega(P)\rightarrow\Omega(Q)$ such that 
$\forall A \in \Omega(P).\ A\mathcal{=}_{pre}\phi(A)$.
\end{definition}

\begin{definition}
Protocol $P$ is a \textit{suffix} of protocol $Q$ ($P$[$\mathcal{S}$]$Q$) if there exists a bijective function $\phi:\Omega(P)\rightarrow\Omega(Q)$ such that 
$\forall A \in \Omega(P).\ A\mathcal{=}_{suf}\phi(A)$.
\end{definition}

\begin{definition}
Protocol $P$ is an \textit{infix} of protocol $Q$ ($P$[$\mathcal{I}$]$Q$) if there exists a bijective function $\phi:\Omega(P)\rightarrow\Omega(Q)$ such that 
$\forall A \in \Omega(P).\ A\mathcal{=}_{inf}\phi(A)$.
\end{definition}

\begin{definition}
Protocol $P$ is a \textit{complement\_to\_infix} of protocol $Q$ ($P$[$\mathcal{C}$]$Q$) if there exists a bijective function $\phi:\Omega(P)\rightarrow\Omega(Q)$ such that 
$\forall A \in \Omega(P).\ A\mathcal{=}_{c}\phi(A)$.
\end{definition}

We denote the composition of relations by sequencing the names of the relations. For instance, the relationship $P[\mathcal{ZRP}]Q$ means that protocol $P$ is a specialization of a restriction of a prefix of protocol $Q$. 

Following, we present first a list of relevant properties regarding the previous relationships and then some proofs of those properties. \\

\subsubsection{Properties of the protocol relationships}

Let $P$ and $Q$ be two protocols. Then, the following properties can be highlighted:

\begin{enumerate}
\item 
$\mathcal{P}$, $\mathcal{S}$, $\mathcal{I}$,  $\mathcal{C}$, $\mathcal{Z}$ and 
$\mathcal{E}$ are reflexive.
\item 
$\mathcal{R}$ is irreflexive.
\item 
$\mathcal{P}$, $\mathcal{S}$, $\mathcal{I}$,  $\mathcal{C}$, $\mathcal{Z}$, $\mathcal{E}$ and $\mathcal{R}$ are transitive.
\item 
$\forall \mathcal{X} \in \{\mathcal{P}, \mathcal{S}, \mathcal{I}, \mathcal{C}, \mathcal{E}, \mathcal{Z}\}.\ [\mathcal{X}\mathcal{R}] = [\mathcal{R}\mathcal{X}]$.
\item 
$\forall \mathcal{X} \in \{\mathcal{P}, \mathcal{S}, \mathcal{I},
 \mathcal{C}\}.\  \forall \mathcal{Y} \in \{\mathcal{E}, \mathcal{Z}\}.\ 
[\mathcal{Y}\mathcal{X}] \Rightarrow [\mathcal{X}\mathcal{Y}]$ but  
 $[\mathcal{X}\mathcal{Y}] \nRightarrow [\mathcal{Y}\mathcal{X}]$.
\item 
$\forall \mathcal{X} \in \{\mathcal{P}, \mathcal{S}, \mathcal{I}\}.\ \forall \mathcal{Y} \in \{\mathcal{P}, \mathcal{S}, \mathcal{I}\}.\ (\mathcal{X}\neq\mathcal{Y} \rightarrow\ (P[\mathcal{XY}]Q \Leftrightarrow\ P[\mathcal{I}]Q)$).
\end{enumerate}

\subsubsection{Proofs}
Next, we provide proofs for some of the properties listed above. Remaining proofs can be 
figured out accordingly. 
\begin{property}:  
$\mathcal{P}$, $\mathcal{S}$, $\mathcal{I}$,  $\mathcal{C}$, $\mathcal{Z}$ and 
$\mathcal{E}$ are reflexive.

\begin{itemize}
\item
$\mathcal{P}$ is reflexive.

\textit{Proof:} We need to prove that $P$[$\mathcal{P}$]$P$.\\
It suffices to define $\phi$ as the identity function, and obviously, $\forall A \in\Omega(P).\ A=_{pre}A$, because $A=_{b}Prune[A/s_{n}]$, where $s_{n}$ is the final state of $A$.\\

\item 
$\mathcal{Z}$ is reflexive.

\textit{Proof:} We need to prove that $P$[$\mathcal{Z}$]$P$.\\
It suffices to define $\phi$ as the identity function, and obviously $\forall A \in\Omega(P),\ A\ll_{b}A$, because for the identity function $id:\mathcal{T}(A)\rightarrow\mathcal{T}(A)$, $\forall t \in\mathcal{T}(A).\ msc(t)\sqsubseteq msc(id(t))$.
\end{itemize}
\end{property}

\begin{property}:
$\mathcal{R}$ is irreflexive.
\begin{itemize}
\item 
\textit{Proof:} We need to prove that $P[\mathcal{R}]P$ is false.
The set $\Omega(P)$ is finite, therefore it is impossible to define a bijective function from $\Omega(P)$ to a proper subset $\Omega'(P)\subset\Omega(P)$.
\end{itemize}
\end{property}
\begin{property}:
$\mathcal{P}$, $\mathcal{S}$, $\mathcal{I}$,  $\mathcal{C}$, $\mathcal{Z}$, $\mathcal{E}$ and $\mathcal{R}$ are transitive.

\begin{itemize}
\item 
$\mathcal{Z}$ is transitive.

\textit{Proof:} We need to prove that $P[\mathcal{ZZ}]Q$ $\Rightarrow$ $P[\mathcal{Z}]Q$. 

$P[\mathcal{ZZ}]Q$ means that there exists a protocol $M$ such that $P[\mathcal{Z}]M$ and $M[\mathcal{Z}]Q$. Looking at the definition of $\mathcal{Z}$, $M[\mathcal{Z}]Q$ means that there exists a bijective function $\phi:\Omega(M)\rightarrow\Omega$(Q) such that 
$\forall A \in \Omega(M).\ A\ll_{b}\phi(A)$. Moreover, $P[\mathcal{Z}]M$ means that 
there exists a bijective function $\phi':\Omega(P)\rightarrow\Omega(M)$ such that 
$\forall A \in \Omega(P).\ A\ll_{b}\phi'(A)$. 
Then, the bijective function $\phi \circ \phi':\Omega(P)\rightarrow\Omega(Q)$ satisfies  
the needed properties since $\forall A \in \Omega(P)\ A\ll_{b}\phi'(A)\ll_{b}\phi \circ \phi'(A)$ 
and $\ll_{b}$ is transitive.\\

\item
$\mathcal{P}$ is transitive.

\textit{Proof:} We need to prove that $P[\mathcal{PP}]Q$ $\Rightarrow$ $P[\mathcal{P}]Q$. 

$P[\mathcal{PP}]Q$ means that there exists a protocol $M$ such that $P[\mathcal{P}]M$ and $M[\mathcal{P}]Q$. Looking at the definition of $\mathcal{P}$, $M[\mathcal{P}]Q$ means that there exists a bijective function $\phi:\Omega(M)\rightarrow\Omega(Q)$ such that 
$\forall A \in \Omega(M).\ \exists sq_{i} \in Q.\ A\mathcal{=}_{b}$\textit{Prune}[$\phi(A)/sq_{i}$]. 
Moreover, $P[\mathcal{P}]M$ means that there exists a bijective function $\phi':\Omega(P)\rightarrow\Omega(M)$ such that 
$\forall B \in \Omega(P).\ \exists sm_{i} \in M.\  B\mathcal{=}_{b}$\textit{Prune}[$\phi'(B)/sm_{i}$]. 
We ask for a small abuse of notation, denoting $\phi(sm_{i})$ to the state in $Q$ corresponding 
to $sm_{i}$ after applying $\phi$ to branch $\phi'(B)$.
Then, the bijective function $\phi \circ \phi':\Omega(P)\rightarrow\Omega(Q)$ is such that $\forall B \in\Omega(P).\ \exists \phi(sm_{i}) \in Q.\ B\mathcal{=}_{b}\textit{Prune}[\phi \circ \phi'(B)/\phi(sm_{i})]$.\\

\item 
$\mathcal{R}$ is transitive.

\textit{Proof:} We need to prove that $P[\mathcal{RR}]Q$ $\Rightarrow$ $P[\mathcal{R}]Q$. 

$P[\mathcal{RR}]Q$ means that there exists a protocol $M$ such that $P[\mathcal{R}]M$ and $M[\mathcal{R}]Q$. Looking at the definition of $\mathcal{R}$, $M[\mathcal{R}]Q$ means that there exists a bijective function $\phi:\Omega(M)\rightarrow\Omega'(Q)$ such that 
$\forall A \in \Omega(M).\ A=_{b}\phi(A)$. Moreover, $P[\mathcal{R}]M$ means that there exists a bijective function $\phi':\Omega(P)\rightarrow\Omega'(M)$ such that 
$\forall B \in \Omega(P)\ B=_{b}\phi'(B)$. Then, the bijective function $\phi \circ \phi':\Omega(P)\rightarrow\Omega''(Q)\subset\Omega(Q)$, where 
$\Omega''(Q)=\phi \circ \phi'(\Omega(P))=\phi(\Omega'(M))$, is such that 
$\forall A \in\Omega(P).\  A\mathcal{=}_{b}\phi \circ \phi'(A)$.

\end{itemize}
\end{property}
\begin{property}:
$\forall \mathcal{X} \in \{\mathcal{P}, \mathcal{S}, \mathcal{I}, \mathcal{C}, \mathcal{E}, \mathcal{Z}\}$. $[\mathcal{X}\mathcal{R}] = [\mathcal{R}\mathcal{X}]$.

\begin{itemize}
\item 
$[\mathcal{P}\mathcal{R}] = [\mathcal{R}\mathcal{P}]$.

\textit{Proof:} First, we prove $P[\mathcal{PR}]Q$ $\Rightarrow$ $P[\mathcal{RP}]Q$.

Because of $P[\mathcal{PR}]Q$, there exists a protocol $M$ such that $P[\mathcal{P}]M$ and $M[\mathcal{R}]Q$. According to definitions of $\mathcal{P}$ and $\mathcal{R}$, 
$\exists\phi: \Omega(P)\rightarrow\Omega(M)$ bijective, such that $\forall A\in\Omega(P).$ $A=_{pre}\phi(A)$ and $\exists\phi': \Omega(M)\rightarrow\Omega'(Q)\subset\Omega(Q)$ 
bijective, such that $\forall A\in\Omega(M).$ $A=_{b}\phi'$(A).

We must prove that there exists a protocol $M'$ such that $P[\mathcal{R}]M'$ and $M'[\mathcal{P}]Q$. Let us define $\Omega$($\overline{M}$) as the set of branches of 
protocol $Q$ that do not have a corresponding branch in $M$  
($\Omega(\overline{M}) = \Omega(Q) - \Omega'(Q)$, notice that 
$\Omega(\overline{M})\neq \emptyset$). Then, $M'$ is a protocol with 
 $\Omega(M') = \Omega(P) \cup \Omega(\overline{M})$.

In order to prove $M'[\mathcal{P}]Q$ we define the bijective function 
$\psi: \Omega(M')\rightarrow \Omega(Q)$ 
\begin{itemize}
\item 
if $A \in \Omega(P)$ then $\psi(A) = \phi'(\phi(A))$.
\item 
if $A \notin \Omega(P)$ then $\psi(A) = A$.
\end{itemize}
Then, $\forall A\in \Omega(M').$ $A=_{pre}\psi(A)$, since 
$A=_{pre}\phi(A)=_{b}\phi'(\phi(A))$ implies $A=_{pre}\phi'(\phi(A))$.

To prove $P[\mathcal{R}]M'$, we use the identity function 
$id: \Omega(P)\rightarrow \Omega(P)\subset \Omega(M')$ so that 
$\forall A\in \Omega(P).$ $A=_{b}id(A)$.\\ 

Second, we prove $P[\mathcal{RP}]Q$ $\Rightarrow$ $P[\mathcal{PR}]Q$.

Because of $P[\mathcal{RP}]Q$, there exists a protocol $M$ such that $P[\mathcal{R}]M$ and $M[\mathcal{P}]Q$. According to definitions of $\mathcal{R}$ and $\mathcal{P}$,
$\exists\phi': \Omega(P)\rightarrow \Omega'(M)\subset\Omega(M)$ bijective, such that 
$\forall A\in \Omega(P).\ A=_{b}\phi'(A)$ and 
$\exists\phi: \Omega(M)\rightarrow \Omega(Q)$ bijective, such that 
$\forall A\in \Omega(M).\ A=_{pre}\phi(A)$.

We must prove that there exists a protocol $M'$ such that $P[\mathcal{P}]M'$ and $M'[\mathcal{R}]Q$.

We build a protocol $M'$ in such a way that 
$\Omega(M') = \{\phi(\phi'(A))\in \Omega(Q) \mid  A \in \Omega(P)\}$. 
In order to prove $P[\mathcal{P}]M'$, we define the function 
$\psi:\Omega(P)\rightarrow \Omega(M')$ such that $\psi(A) = \phi(\phi'(A)$. 
Then, $\psi$ is bijective and 
$\forall A\in\Omega(P).\ A =_{b} \phi'(A) =_{pre} \phi(\phi'(A)$, and consequently 
$A =_{pre}\psi(A)$.

Moreover, the identity function 
$id:\Omega(M')\rightarrow \Omega(M')\subset \Omega(Q)$  is bijective and satisfies the condition 
for $M'[\mathcal{R}]Q$, by construction of $\Omega(M')$.\\

\item 
$[\mathcal{Z}\mathcal{R}] = [\mathcal{R}\mathcal{Z}]$.

\textit{Proof:} First, we prove $P[\mathcal{ZR}]Q$ $\Rightarrow$ $P[\mathcal{RZ}]Q$. 

Because of $P[\mathcal{ZR}]Q$, there exists a protocol $M$ such that $P[\mathcal{Z}]M$ and $M[\mathcal{R}]Q$. 
According to definitions of $\mathcal{Z}$ and $\mathcal{R}$, 
$\exists\phi:\Omega(P)\rightarrow\Omega(M)$ bijective, such that 
$\forall A \in \Omega(P).\ A\ll_{b}\phi(A)$ 
 and $\exists\phi': \Omega(M)\rightarrow\Omega'(Q)\subset\Omega(Q)$ 
bijective, such that $\forall A\in\Omega(M).\ A=_{b}\phi'$(A).

We must prove that there exists a protocol $M'$ such that $P[\mathcal{R}]M'$ and $M'[\mathcal{Z}]Q$.

Let us define $\Omega$($\overline{M}$) as the set of branches of 
protocol $Q$ that do not have a corresponding branch in $M$  
($\Omega(\overline{M}) = \Omega(Q) - \Omega'(Q)$). Then,  $M'$ is a protocol with 
 $\Omega(M') = \Omega(P) \cup \Omega(\overline{M})$.

In order to prove $M'[\mathcal{Z}]Q$ we define the bijective function 
$\psi: \Omega(M')\rightarrow \Omega(Q)$ 
\begin{itemize}
\item 
if $A \in \Omega(P)$ then $\psi(A) = \phi'(\phi(A))$.
\item 
if $A \notin \Omega(P)$ then $\psi(A) = A$.
\end{itemize}
Then, $\forall A\in \Omega(M').$ $A\ll_{b}\psi(A)$, since 
$A\ll_{b}\phi(A)=_{b}\phi'(\phi(A))$ implies $A\ll_{b}\phi'(\phi(A))$.

Moreover, $P[\mathcal{R}]M'$ because the identity function 
$id:\Omega(P)\rightarrow \Omega(P)\subset \Omega(M')$ is bijective and satisfies 
the needed properties, by construction of $\Omega(M')$.

Second, we prove $P[\mathcal{RZ}]Q$ $\Rightarrow$ $P[\mathcal{ZR}]Q$. 

Because of $P[\mathcal{RZ}]Q$, there exists a protocol $M$ such that $P[\mathcal{R}]M$ 
and $M[\mathcal{Z}]Q$. 
According to definitions of $\mathcal{R}$ and $\mathcal{Z}$, 
$\exists\phi:\Omega(P)\rightarrow\Omega'(M)\subset \Omega(M)$ bijective, such that 
$\forall A \in \Omega(P).\ A=_{b}\phi(A)$ 
 and $\exists\phi': \Omega(M)\rightarrow\Omega(Q)$ 
bijective, such that $\forall A\in\Omega(M).\ A\ll_{b}\phi'$(A).

We must prove that there exists a protocol $M'$ such that $P[\mathcal{Z}]M'$ and $M'[\mathcal{R}]Q$.

 Then we build a protocol $M'$ with the branches of $Q$ that have a corresponding branch 
in $\Omega(M)$ due to $\phi'$ ($\Omega(M')=\phi'(\Omega'(M)))$. 

In order to prove  $P[\mathcal{Z}]M'$ we define the bijective function 
$\psi: \Omega(P)\rightarrow \Omega(M')$ with $\psi(A) = \phi'(\phi(A))$.
Then, $\forall A\in \Omega(P).$ $A\ll_{b}\psi(A)$, since 
$A=_{b}\phi(A)\ll_{b}\phi'(\phi(A))$ implies $A\ll_{b}\phi'(\phi(A))$.

Finally, $M'[\mathcal{R}]Q$ because the identity function 
$id:\Omega(M')\rightarrow \Omega(M')\subset \Omega(Q)$ is bijective and satisfies 
the needed properties, by construction of $\Omega(M')$.

\end{itemize}
\end{property}

\begin{property}:
$\forall \mathcal{X} \in \{\mathcal{P}, \mathcal{S}, \mathcal{I},
 \mathcal{C}\}.\  \forall \mathcal{Y} \in \{\mathcal{E}, \mathcal{Z}\}.\ 
[\mathcal{Y}\mathcal{X}] \Rightarrow [\mathcal{X}\mathcal{Y}]$ but  
 $[\mathcal{X}\mathcal{Y}] \nRightarrow [\mathcal{Y}\mathcal{X}]$.

\begin{itemize}
\item 
$P[\mathcal{ZP}]Q$ $\Rightarrow$ $P[\mathcal{PZ}]Q$.

\textit{Proof:} 
Because of $P[\mathcal{ZP}]Q$, there exists a protocol $M$ such that $P[\mathcal{Z}]M$ and $M[\mathcal{P}]Q$. According to the definitions of $\mathcal{Z}$ and $\mathcal{P}$, $\exists\phi:\Omega(P)\rightarrow\Omega(M)$ bijective, such that 
$\forall A \in \Omega(P).\ A\ll_{b}\phi(A)$ 
 and $\exists\phi': \Omega(M)\rightarrow\Omega(Q)$ 
bijective, such that $\forall B\in\Omega(M).\ B=_{pre}\phi'(B)$, which implies that 
$\forall B\in\Omega(M)$. $\exists s_{\phi'(B)}.\ B=_{b} Prune[\phi'(B)/s_{\phi'(B)}]$ and 
exists \textit{Suff}($\phi'(B)$)=\textit{ChangeInit}[$\phi'(B)|s_{\phi'(B)}$] 
(In other words, each branch of $Q$ can be seen as a concatenation of two subbranches: 
$B\cdot$\textit{Suff}($\phi'(B)$)).
 Let us define $\Omega(\overline{M}) = \{\textit{Suff}(\phi'(B))| B \in \Omega(M)\}$.

 We must prove that there exists a protocol $M'$ such that $P[\mathcal{P}]M'$ and $M'[\mathcal{Z}]Q$.

Then we build a protocol $M'$ with the branches formed by the concatenation of each branch in 
 $\Omega(P)$ with its corresponding branch in $\Omega(\overline{M})$ 
($\Omega(M') = \{A\cdot\textit{Suff}(\phi'\circ\phi(A))| A\in \Omega(P)\}$). 

In order to prove $P[\mathcal{P}]M'$ we define the bijective function 
$\psi:\Omega(P)\rightarrow \Omega(M')$ with $\psi(A) = A\cdot\textit{Suff}(\phi'\circ\phi(A))$, 
which satisfies $A=_{pre}\psi(A)$ by construction.

In order to prove  $M'[\mathcal{Z}]Q$, notice that every $B\in \Omega(M')$ is of the 
form $A\cdot\textit{Suff}(\phi'\circ\phi(A))$ for some $A\in \Omega(P)$; then  we define 
the bijective function $\psi:\Omega(M')\rightarrow \Omega(Q)$ with 
$\psi(A\cdot\textit{Suff}(\phi'\circ\phi(A))) = \phi(A)\cdot\textit{Suff}(\phi'\circ\phi(A))$, 
which satisfies 
$A\cdot\textit{Suff}(\phi'\circ\phi(A))\ll_{b}\phi(A)\cdot\textit{Suff}(\phi'\circ\phi(A))$ 
by definition of $\phi$. \\

\item 
$P[\mathcal{PZ}]Q$ $\nRightarrow$ $P[\mathcal{ZP}]Q$.

We show a counter-example.
\begin{figure*}
	\centering
	\includegraphics[width=2.0in]{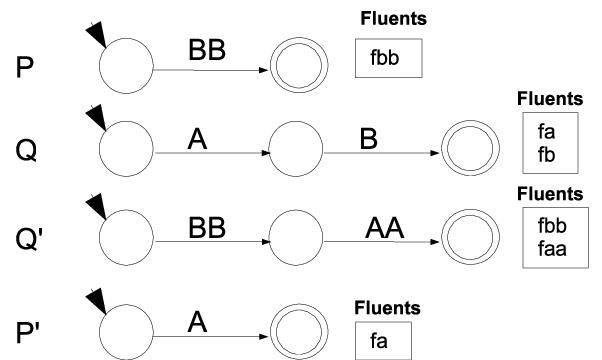}
	\caption{Counter-example: $\mathcal{PZ}\nRightarrow\mathcal{ZP}$}
	\label{PZnotZP}
\end{figure*}

Because of $P[\mathcal{PZ}]Q$, there exists a protocol $M$ such that $P[\mathcal{P}]M$ and 
$M[\mathcal{Z}]Q$. Looking at Fig.~\ref{PZnotZP}, let us suppose that 
$msc(fbb)\sqsubseteq msc(fb)$, $msc(faa)\sqsubseteq msc(fa)$ and no other relationship exists 
between fluents, so $Q'$ in the picture fulfills the requirements. 

We must prove that there does not exist a protocol $M'$ such that $P[\mathcal{Z}]M'$ 
and $M'[\mathcal{P}]Q$. 
Notice that any protocol $M'$ such that $M'[\mathcal{P}]Q$, should satisfy 
$M'[\mathcal{E}]Q$ or $M'[\mathcal{E}]Q'$; but it is obvious that 
not $P[\mathcal{Z}]Q$ neither $P[\mathcal{Z}]Q'$.


\end{itemize}
\end{property}

\begin{property}:
$\forall \mathcal{X} \in \{\mathcal{P}, \mathcal{S}, \mathcal{I}\}.\ \forall \mathcal{Y} \in \{\mathcal{P}, \mathcal{S}, \mathcal{I}\}.\ (\mathcal{X}\neq\mathcal{Y} \rightarrow\ (P[\mathcal{XY}]Q \Leftrightarrow\ P[\mathcal{I}]Q)$).

%
%
\begin{itemize}
\item $P[\mathcal{PS}]Q$ $\Leftrightarrow$ $P[\mathcal{I}]Q$. (A prefix of a suffix is an infix, and viceversa).

\textit{Proof}: First we prove $[\mathcal{PS}]\Rightarrow[\mathcal{I}]$.

Because of $P[\mathcal{PS}]Q$, there exists a protocol $M$ such that $P[\mathcal{P}]M$ and $M[\mathcal{S}]Q$. According to the definitions of $\mathcal{P}$ and $\mathcal{S}$, $\exists\phi:\Omega(P)\rightarrow\Omega(M)$ bijective, such that 
$\forall A \in \Omega(P).\ A=_{pre}\phi(A)$ 
 and $\exists\phi': \Omega(M)\rightarrow\Omega(Q)$ 
bijective, such that $\forall B\in\Omega(M).\ B=_{suf}\phi'$(B). 

Then, the bijective function $\psi:\Omega(P)\rightarrow\Omega(Q)$ with 
$\psi(A) = \phi'(\phi(A))$ satisfies $A=_{inf}\psi(A)$, 
since $\forall A \in \Omega(P).\ A=_{pre}\phi(A)=_{suf}\phi'(\phi(A))$, 
and due to definition of $=_{suf}$, 
there exists a state $s_{a}$ in $\phi'(\phi(A))$ such that 
$A=_{pre}\phi(A)=_{b}\textit{ChangeInit}[\phi'(\phi(A))/s_{a}]$, 
which implies that 
$A=_{pre}\textit{ChangeInit}[\phi'(\phi(A))/s_{a}]$, 
which is the definition for $P[\mathcal{I}]Q$.

Next we prove $[\mathcal{I}]\Rightarrow[\mathcal{PS}]$.

Because of $P[\mathcal{I}]Q$, $\exists\phi:\Omega(P)\rightarrow\Omega(Q)$ bijective, such that $\forall A \in \Omega(P).\ A=_{inf}\phi(A)$, which means that $\forall A\in\Omega(P)$ there exists a state $s_{\phi(A)}$ in $\phi(A)$ such that ${A}\mathcal{=}_{pre}$\textit{ChangeInit}[$\phi(A)$/$s_{\phi(A)}]=_{suf}\phi(A)$.
We must prove that there exists a protocol $M$ such that $P[\mathcal{P}]M$ and $M[\mathcal{S}]Q$. We define $\Omega(M)=\{\textit{ChangeInit}[\phi(A)/s_{\phi(A)}]\mid\ A\in\Omega(P)\}$. 
Then, the functions $\psi:\Omega(P)\rightarrow\Omega(M)$ with 
 $\psi(A)=\textit{ChangeInit}[\phi(A)/s_{\phi(A)}]$ and 
$\psi':\Omega(M)\rightarrow\Omega(Q)$ with 
$\psi'(\textit{ChangeInit}[\phi(A)/s_{\phi(A)}]) = \phi(A)$ are bijective and justify 
 $P[\mathcal{P}]M$ and $M[\mathcal{S}]Q$.

\end{itemize}

\end{property}

%
%
%

To conclude this section, we want to point out that our implemented mechanism discovers 
complex composition relationships, like those presented in this section, between two 
given protocols. The following section presents an example. 

\section{One scenario of the proposed mechanism at work}\label{Example}
The aim of this section is to show one scenario that illustrates the different steps that the proposed mechanism follows in order to discover the relationship between two different protocols that could be activated in case of a road accident. In both cases, the actors that interact are a nurse from the ambulance that covers the emergency and a doctor from the emergency staff from a hospital. Those actors are represented by software agents that belong to different Information Systems. Notice that those agents use protocols constituted by communication acts that are specific to each Information System, and which are a specialization of general communication act classes defined in the upper level of the \textsc{CommOnt} ontology.

\begin{figure}
	\centering
	\includegraphics[width=5.0in]{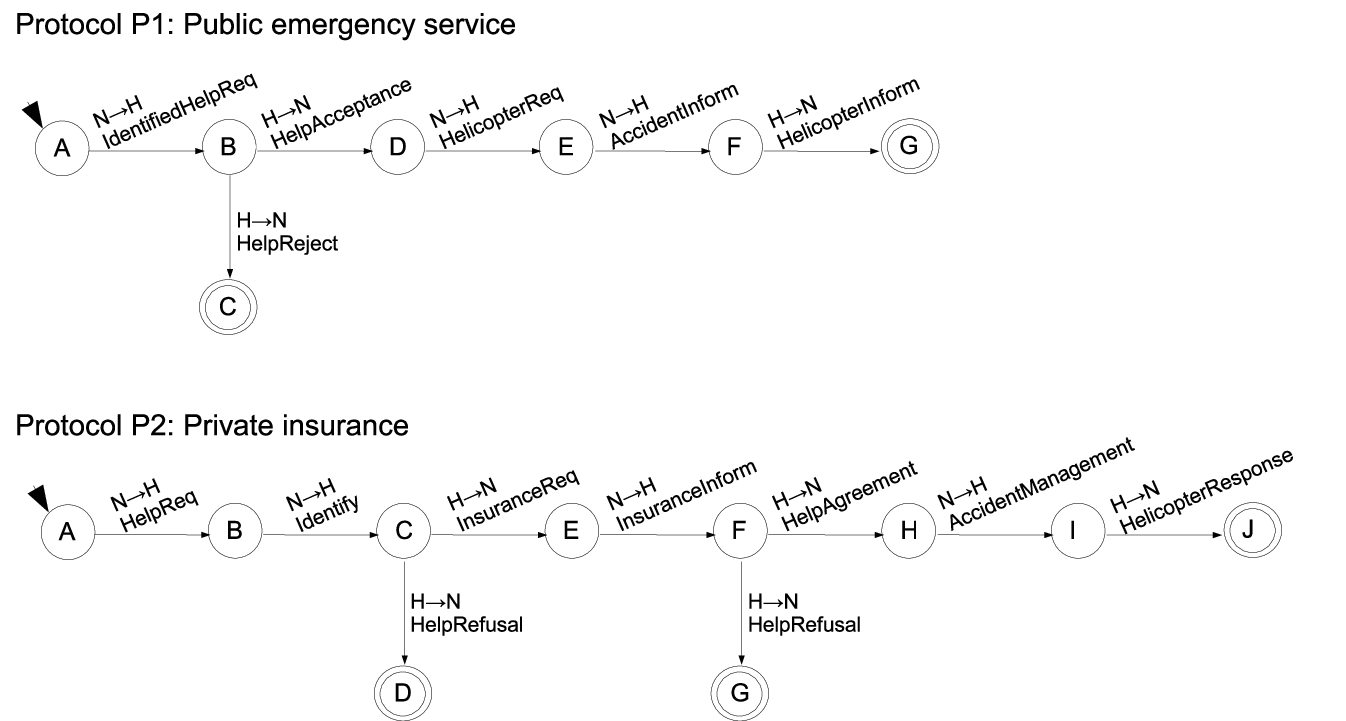}
	\caption{One scenario of the proposed mechanism.}
\label{protocolsExample}
\end{figure}

On the one hand, protocol $P1$ in Fig.~\ref{protocolsExample} illustrates the interaction with a public hospital. The protocol is initiated by a nurse from the ambulance, who sends a request (\texttt{IdentifiedHelpReq}) for help identifying herself within the message. Depending on the situation, the hospital staff can either reject (\texttt{HelpReject}) or accept the request and ask information about the accident (\texttt{HelpAcceptance}). In the latter case, then the nurse requests a helicopter to take the injured to the hospital (\texttt{HelicopterReq}) and sends a message with the information about the accident (\texttt{AccidentInform}). Finally, the hospital staff responds whether it is possible or not to send a helicoper (\texttt{HelicopterInform}).

On the other hand, protocol $P2$ shows the interaction with a private hospital, where patients are required to have an insurance policy. In the first step, the nurse sends a request for help (\texttt{HelpReq}) followed by a message of self-identification (\texttt{Identify}). The hospital staff may either reject the request (\texttt{HelpRefusal}) or accept it (\texttt{InsuranceReq}), in which case the nurse informs about the patient's insurance details (\texttt{InsuranceInform}). After checking those details, the hospital staff decides whether the patient is eligible for help (\texttt{HelpAgreement}) or not (\texttt{HelpRefusal}). In the former case, then the nurse sends a message in which she gives information about the accident and requests a helicopter (\texttt{AccidentManagement}). Finally, the hospital staff responds whether it is possible or not to send a helicoper (\texttt{HelicopterResponse}).

\begin{figure}
	\centering
	\includegraphics[width=6.0in]{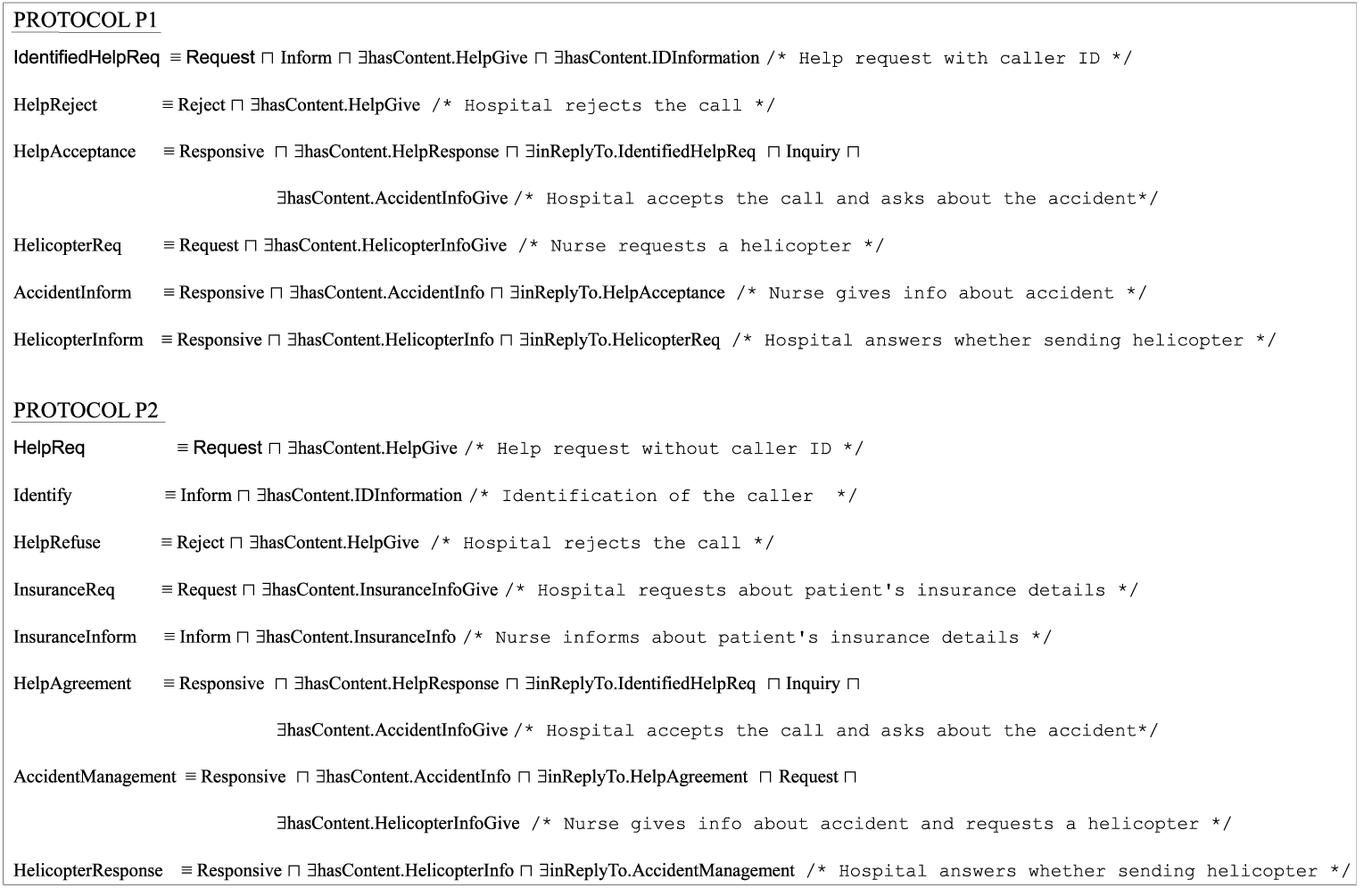}
	\caption{Description of the communication acts}
	\label{prots1and2CommActs}
\end{figure}

It can be noticed that both protocols are quite different. In Fig.~\ref{prots1and2CommActs} the description and explanations of the communication acts that appear in the protocols are provided.

In the following, the different steps presented in the previous sections (\ref{Separation}, \ref{Generation} and \ref{Pairing}) are illustrated for the scenario under consideration.\\

\underline{Separation into branches:}
There are two different branches in protocol $P1$ and three different branches in protocol $P2$. For brevity, we represent a branch with the sequence of its states. Then, protocol $P1$ is separated into the branches $B1.1$=[A,B,C] and $B1.2$=[A,B,D,E,F,G] and protocol $P2$ into $B2.1$=[A,B,C,D], $B2.2$=[A,B,C,E,F,G] and $B2.3$=[A,B,C,E,F,H,I,J].\\

\begin{figure}
	\centering
	\includegraphics[width=5.5in]{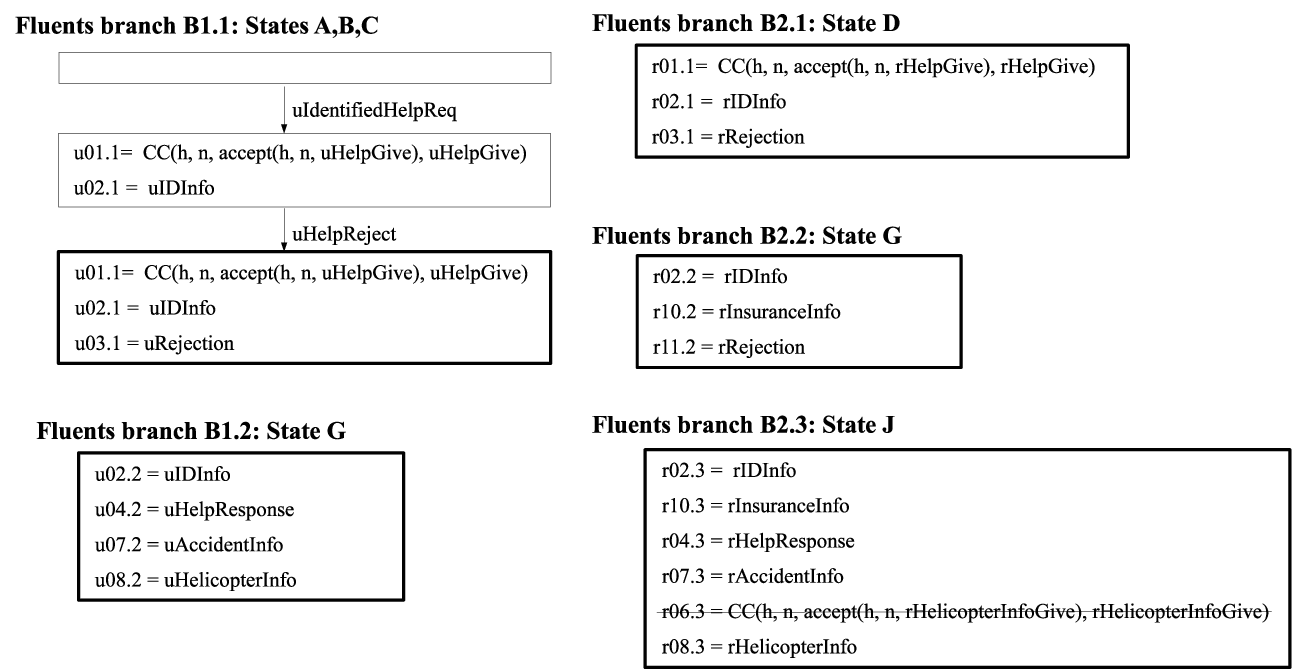}
	\caption{Fluents generated by the branches}
	\label{lastStatesFluents}
\end{figure}

\underline{Generation of fluents:}
In Fig.~\ref{lastStatesFluents} the fluents generated by each branch are shown. The fluents generated by protocol $P1$ and the instances of communication acts in that protocol are prefixed by \textit{u} (p\underline{u}blic), while the ones generated by protocol $P2$ are prefixed by \textit{r} (p\underline{r}ivate). Below a detailed description of the generation process of the fluents in branch $B1.1$ is provided. The generation process of the remaining branches can be found in appendix~\ref{GenerationOfFluents}.\\

\textit{Branch B1.1}\\

\underline{$<A, IdentifiedHelpReq, B>$:} Nurse \textit{n} sends the message \textit{uIdentifiedHelpReq}\\

Due to the following knowledge:
\scriptsize

\begin{center}
\texttt{uIdentifiedHelpReq} $\in$  \texttt{IdentifiedHelpReq}\\
\texttt{IdentifiedHelpReq} $\sqsubseteq$ \texttt{Request} \\
\texttt{IdentifiedHelpReq} $\sqsubseteq$ \texttt{Inform} $\sqsubseteq$  \texttt{Assertive} 
\end{center}
\begin{center} \textit{Initiates(Request(s,r,P)}, \textsf{CC}\textit{(r, s, accept(r,s,P), P), t)} \\
\textit{Initiates(Assertive(s,r,P), {P}, t)} \end{center}
\normalsize
fluents $u01.1$ and $u02.1$ are generated.\\

\underline{$<B, HelpReject, C>$:} Hospital \textit{h} sends the message \textit{uHelpReject}.\\

Due to the following knowledge:
\scriptsize
\begin{center}
\texttt{uHelpReject} $\in$  \texttt{HelpReject} \\
\texttt{HelpReject} $\sqsubseteq$ \texttt{Reject}
 \end{center}
\begin{center}\textit{Initiates(Reject(s,r,P), {Rejection(s,r,P)}, t)} \end{center}
\normalsize

fluent $u03.1$ is generated.\\

\underline{Comparison of branches:}
\begin{figure}
	\centering
	\includegraphics[width=3.0in]{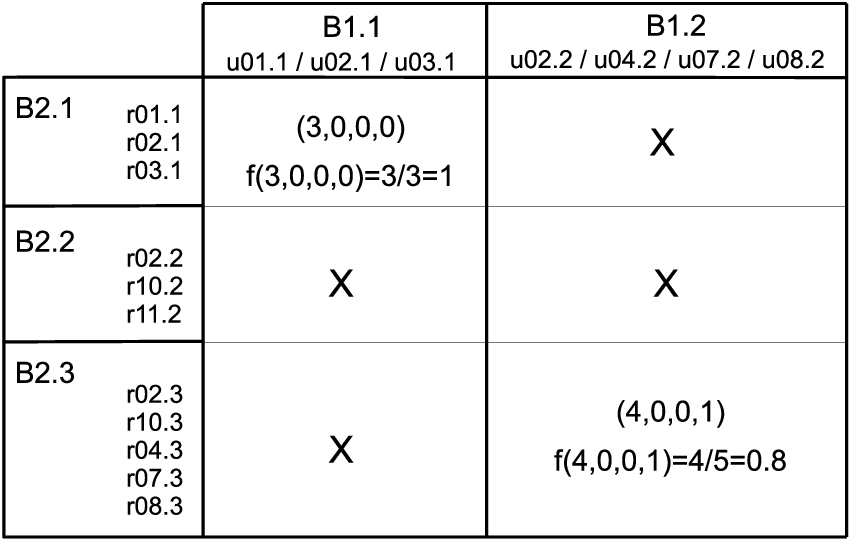}
	\caption{\label{comparisonTable}Fluent comparison table}
\end{figure}
For the sake of the example, we have re-labelled the generated fluents with a code in the format \textit{pX.Y}, where \textit{p} $\in$ \{u,r\} refers to the protocol that generates the fluent, \textit{X} $\in$ \{01,...,n\} is a key number for identifying fluents and \textit{Y} $\in$ \{1,2,3\} is a number for distinguising fluents from diferent branches.
Notice that in our coding, number \textit{X} identifies a fluent, that is to say 
$r02.1$ is the same individual as $u02.1$, their string labels is the sole 
difference.
Taking the previous considerations into account, the branches comparison table displays the results in Fig.~\ref{comparisonTable}.

Remember that the tuple (4,0,0,1) indicates that four fluents in the trace of $B2.3$ and four fluents in the trace of $B1.2$ can be paired up through an equivalence relationship, 
no one fluent can be paired up through a specialization relationship and there is one exceeding fluent (due to the fact that the trace of $B2.3$ has five fluents and the trace of $B1.2$ has only four fluents).

Moreover, there are four infeasible pairs  
($B1.2$, $B2.1$), ($B1.1$, $B2.2$), ($B1.2$, $B2.2$) and 
($B1.1$, $B2.3$). For example, ($B1.1$, $B2.2$) is not feasible because in the trace of $B1.1$ there are some fluents ($u01.1$ and $u03.1$) which do not hold in the trace of $B2.2$, and in turn, in trace of $B2.2$ there are other fluents ($r10.2$ and $r11.2$) which do not hold in trace of $B1.1$.

The best pairing can be obtained pairing up branch $B1.1$ with branch $B2.1$ and branch $B1.2$ with $B2.3$ ($\pi = \{(B1.1, B2.1), (B1.2, B2.3)\}$). In addition, since all the fluents in $B1.1$ have been related through an equivalence relationship \textit{(eq)} with those in $B2.1$ ((3,0,0,0) in the table) it can be said that those branches are equivalent ($B1.1=_{b}B2.1$). Furthermore, due to the (4,0,0,1) valuation given to ($B1.2$, $B2.3$), it can be seen that $B1.2$ is somehow included in $B2.3$. Since the fluent that has not been paired up ($r10.3$) is generated in an inner state of $B2.3$, it results $B1.2=_{c}B2.3$.

In summary, two of the three branches of $P2$ can be related with the ones in $P1$ (notion of 
restriction $\mathcal{R}$), those branches are related via the complement\_to\_infix 
relationship $\mathcal{C}$ and the relationship between the compared fluents is the one of equivalence $\mathcal{E}$, so we can say that $P1[\mathcal{ECR}]P2$.

\section{Related Works}\label{trabajos} 
Different criteria can be considered for a classification of related works 
that can be found in the specialized literature. One significant criterion, with respect to protocol definitions, 
is that which distinguishes among those works that prevail in a semantic approach in contrast with others that only consider a structural one. 
Another interesting criterion is the consideration or not of formal relationships 
between protocols. 
A third criterion is if the works consider a notion of protocol composition.

We will firstly review the first criterion that is, works which introduce semantic considerations  
when defining protocols. 

The closer work is \citep{MallyaSingh2007}, where protocols are represented as transition 
systems where transitions are formalized as operations on commitments.  
Subsumption and equivalence of protocols are defined with respect to three state similarity 
funtions. We share some goals with that work, but in that paper there are no references 
to how protocol relationships are computed. In contrast, according to our proposal, protocol 
relationships can be computed by straightforward algorithms. It is worth 
mentioning that protocol relationships considered in that paper deserve study within our 
framework. Another work that consider protocol relationships is \citep{DesaiMallyaEtAl2005}, 
where protocols are represented with a set of rules with terms 
obtained from an ontology. In particular they formalize protocols into the $\pi$-calculus; 
then, equivalence through bisimulation is the only relationship considered.

With respect to the group of works that do not consider the study of relationships 
between protocols, the works \citep{YolumSingh2002} and \citep{Fornara03} are quite 
similar one to another. Both capture the semantics of communication acts 
through agents' commitments and represent communication protocols using a set of rules that 
operate on these commitments. Moreover, those rule sets can be compiled as Finite State Machines. 
In \citep{KagalFinin2007}, protocols are defined as a set of permissions and 
obligations of agents participating in the communication. They use an OWL ontology 
for defining the terms of the specification language, but their basic reasoning is 
made with an \textit{ad hoc} reasoning engine. We share their main goal of defining protocols 
in a general framework that allows them to be re-used. 

One work that focuses on the problem of protocol composition is 
\citep{YS07}. It introduces considerations of rationality on the enactment of 
protocols. Our proposal could be complemented with the ideas presented in that work.

We review secondly works that deal solely with structural aspects of 
protocols. We advocate for a more flexible approach dealing with semantics.

The work presented in \citep{Montes06} describes a methodology that facilitates communication 
among heterogeneous negotiation agents based on the alignment of communication primitives. 
Finite State Machines are used as a model to represent negotiation protocols and as a base 
for aligning primitives. This is quite a rigid approach because communication will only be 
possible if protocols have such a similar structure that their communication primitives 
can be aligned two by two.

\citep{D'inverno98} and \citep{Mazouzi02} use Finite State Machines and Petri nets, 
respectively, to represent protocols but without taking into account 
the meaning of the communication acts interchanged or considering relationships 
between protocols. The work of \citep{Ryu07} has an interesting approach to manage changes in business 
protocols. More precisely, it presents an extensive study on how to translate active instances 
from an old protocol to a new one, without violating several types of constraints. 
In order to do so, it compares the old protocol with the new one but only examines structural 
differences between them, which results in a more rigid approach that the one allowed by 
our mechanism, which considers semantics representation of protocols.

For the third criterion we issue the problem of determining if an agent's policy conforms to a protocol. This is a very relevant problem but not one which we are examining in this paper. 
Nevertheless, the topic is closely related to ours and it is worth mentioning here.    
In \citep{Endriss03}, deterministic Finite State Machines are the abstract models for 
protocols, which are described by simple logic-based programs. Three levels of 
conformance are defined: weak, exhaustive and robust.
They consider communication acts as atomic actions, in contrast to our semantic view. 
In \citep{Baldoni06-ICSOC} a nondeterministic Finite State Automata is used to support a 
notion of conformance that guarantees interoperability among agents which conform to a 
protocol. Their conformance notion considers the branching structure of policies and 
protocols and applies a simulation-based test. Communication acts are considered 
atomic actions, without considering their semantics.  In \citep{Chopra-Singh-06}, a notion 
of conformance is defined and, moreover, it is proved orthogonal to their proposed 
notions of coverage and interoperability.

Apart from the previous classification, in the context of Web Services, 
state transition systems are used in 
\citep{BordeauxBerardiEtAl2004} for representing dynamic behaviour of services and 
defining some notions 
of compatibility and substitutability of services that can be easily translated to 
the context of compatibility of protocols. Relationships between their compatibility 
relations and our defined relationships deserve study.

\section{Conclusions}
In this paper we have explained a mechanism for discovering semantic relationships among agent 
communication protocols. The mechanism is based on the idea that different protocols can be 
compared semantically by looking to the set of fluents associated to the branches of 
protocols. That assumption favours a much more flexible comparison of protocols than the more 
traditional one based on comparing protocol structures.
Through the paper we have shown first, one ontology that represents concepts related to 
communication acts that agents use to communicate with each other and which take part of the 
protocols. Secondly, we have concentrated on showing: how protocols are modelled,  
in our case using OWL-DL language, and the features of the algorithm that 
analyses each protocol by decomposing it into different branches, and by generating the 
semantics information associated to each branch. Thirdly, we have presented different definitions 
that are managed by the mechanism which permit the identification of protocol relationships. 
Moreover, some properties of those relationships and proofs for them have been included. 
Finally, with an example we illustrated the feasibility of the proposed mechanism which has been 
implemented using Java as a programming language and Pellet as a description logic reasoner. 
As a future work we are studying the interest of adapting the proposed mechanism to contexts 
where Information Systems are represented through Web Services.


\begin{acknowledgements}
The work of Idoia Berges is supported by a grant of the Basque Government (Programa de Formaci\'on
de Investigadores del Departamento de Educaci\'on, Universidades
e Investigaci\'on). This work is also supported by the Basque Country Government IT-427-07 and the Spanish Ministry of Education and Science TIN2007-68091-C02-01.
\end{acknowledgements}

\bibliographystyle{spbasic}

\bibliography{BDI}
\newpage
\appendix
\section{Generation of fluents}\label{GenerationOfFluents}
\begin{figure}[H]
	\centering
	\includegraphics[width=4.3in]{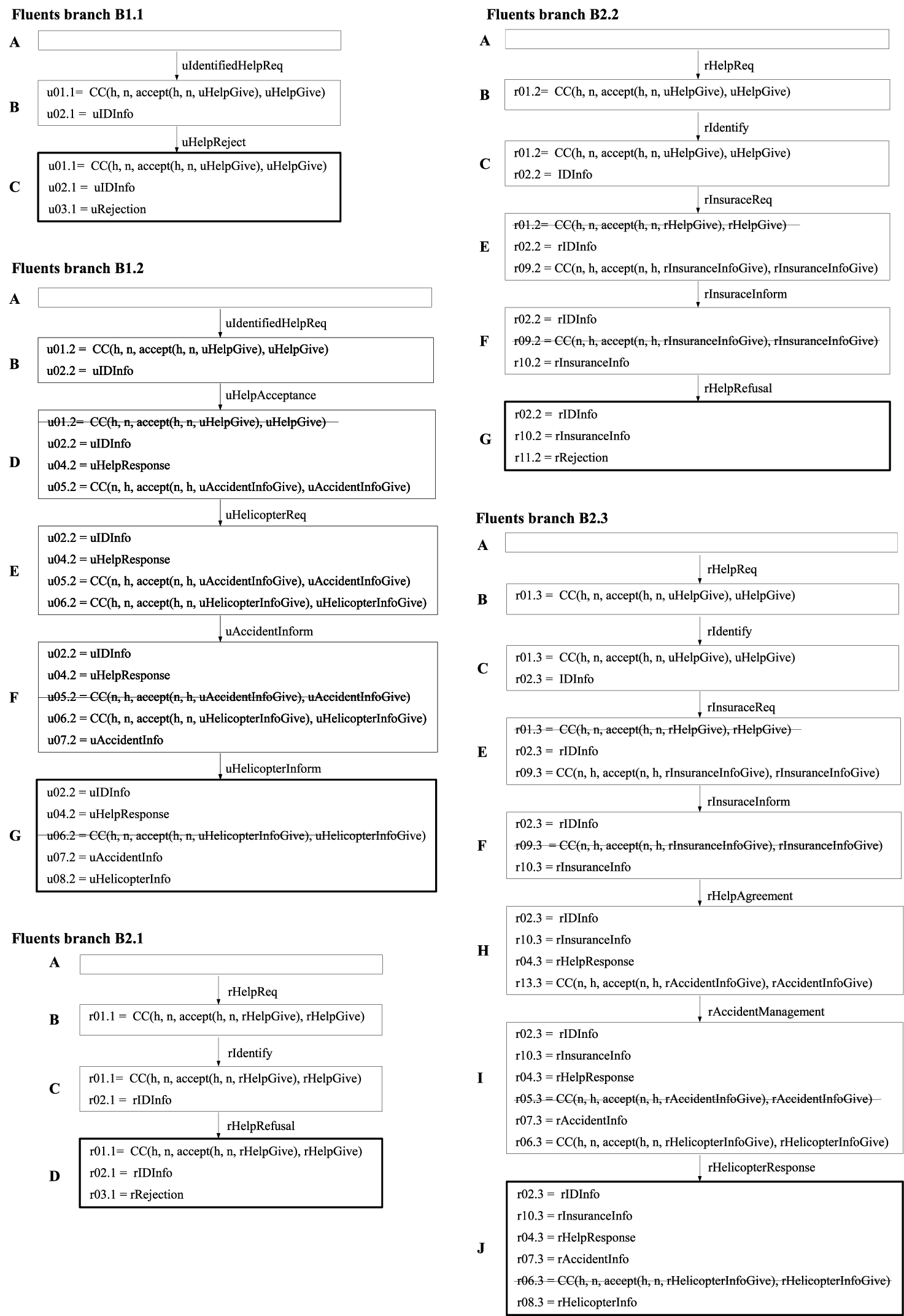}
	\label{prots1and2Fluents}
\end{figure}

%

%


%
%

\end{document}